\documentclass[aps,pra,showpacs,reprint,superscriptaddress]{revtex4-1}
\usepackage{graphicx}
\usepackage{amsmath,amssymb,bm}
\usepackage{siunitx}
\usepackage{subfig}
\usepackage[utf8]{inputenc}
\usepackage[francais,british]{babel}
\usepackage{comment}
\usepackage{color}   
\usepackage{xcolor}
\usepackage{tikz}

\begin{document}

%\preprint{}

\title{Mid- and far-field deviations from causality in spontaneous light emission by atomic Hydrogen}



\author{Vincent Debierre}
\email[Contact:]{vincent.debierre@fresnel.fr}
%\thanks{}
\author{Thomas Durt}
%\email[Edouard.Brainis@UGent.be]{Your e-mail address}
%\thanks{}
\affiliation{Aix Marseille Université, CNRS, École Centrale de Marseille, Institut Fresnel UMR 7249, 13013 Marseille, France.}


\date{\today}

\begin{abstract}
We investigate, in the case of the $2\mathrm{p}-1\mathrm{s}$ transition in atomic Hydrogen, the behaviour of the spontaneously emitted electromagnetic field in spacetime. We focus on Glauber's wave function for the emitted photon, a quantity which we find is nonzero outside the lightcone at all times after the start of the emission. We identify the uncertainty on the position of the decaying electron as a source of departure from causality in the naive sense of the term. We carry out a detailed study of the emitted electric field in the mid- and far-field regions, through analytical and numerical computations as well as asymptotic arguments.
\end{abstract}

%\keywords{Atomic decay, Zeno effect}

\maketitle


\section{Introduction} \label{sec:Intro}

Fermi was the first to study the electromagnetic field emitted during an atomic transition, which has since then been a recurring theme of investigation in atomic physics and quantum electrodyamics \cite{FermiCausal,Shirokov,PassanteVirtual,HegerfeldtFermi,Sipe}. Fermi \cite{FermiCausal} found that, if the survival probability of the electron in the excited state is assumed to decay exponentially, according to the usual Wigner-Weisskopf aproximation \cite{WW}, then the emitted electromagnetic field will propagate causally, in other words, vanish outside the lightcone centred around $t=0$ (the instant at which the emission starts) and the position $\mathbf{x}=\mathbf{0}$ of the Hydrogen nucleus (proton). It was then noticed by Shirokov \cite{Shirokov} that Fermi's result made use of an unmentioned approximation, which consisted in extending the range of integration over electromagnetic frequencies from the positive real semi-axis to the whole real axis, thereby including nonphysical electromagnetic negative frequency modes in the treatment. Hegerfeldt later \cite{HegerfeldtFermi} formalised and generalised Shirokov's remarks, linking the absence of negative electromagnetic frequencies with the noncausal field propagation via the Paley-Wiener theorem \cite{PaleyWiener} on holomorphic Fourier transforms.\\

% The relevance of negative frequencies was further highlighted by the work of Biswas \emph{et al.}, who showed \cite{PassanteVirtual} in a particular case that going beyond the oft-used Rotating Wave Approximation (RWA) led to the natural appearance of ``negative frequencies'' in the computations of the emitted field (or, in their case, and equivalently, the rate of change of the population of the excited atomic level), thereby restoring the causality of light propagation. Their result is further commented upon in sects.~\ref{sec:CausalWW} and \ref{sec:Dsc}.

Note that in \cite{FermiCausal,Shirokov,PassanteVirtual,HegerfeldtFermi} the authors consider not one atom in free space but two, the first one being initially in its excited state while the second one starts in its ground state. Rather than computing the emitted field, they focus on the probability of excitation of the second atom as a function of time (this is known in the literature as the ``Fermi problem''). Our present study features a single atom, initially in its excited state, and purports to obtain the spacetime dependence of the spontaneously emitted field, expressed by Glauber's photon wave function. 

Our main new result is the following: even if we artificially integrate over negative frequencies, thereby bypassing Hegerfeldt's objections, causality is still violated, which establishes that Hegerfeldt's mechanism is not the single source of noncausality in the present problem. This result follows from a rigorous computation of the single photon wave function, and by the use of exact expressions for the coupling coefficients between electronic $1\mathrm{s}$ and $2\mathrm{p}$ states of the Hydrogen atom and electromagnetic modes \cite{Seke,FacchiPhD}.\\

In sect.~\ref{sec:AtomQED} we review the tools for the description of spontaneous emission. Sect.~\ref{sec:GlauberRule} sets the stage for the computation of the wave function of the emitted photon, which is the object which we use in order to assess causality. In sect.~\ref{sec:CausalWW} we review the usual treatment \cite{FermiCausal,FoxMulder} where causality is derived by the way of multiple approximations (dipole approximation, extension of the electromagnetic spectrum to negative frequencies, usual $\hat{\mathbf{E}}\cdot\hat{\mathbf{x}}$ coupling instead of the minimal $\hat{\mathbf{A}}\cdot\hat{\mathbf{p}}$ coupling). In sects.~\ref{sec:PreStorm} and \ref{sec:MyHeart} we refine the treatment by progressively waiving various approximations, and we find that the result is no longer causal. We identify the different sources of noncausality, and study the spacetime dependence of the photon wave function in the mid- and far-field regions in detail. Sect.~\ref{sec:Dsc} is a discussion of our results.

\section{The decay of a two-level atom} \label{sec:AtomQED}

Let us consider a two-level atom (ground state $\mid\!\mathrm{g}\rangle$, excited state $\mid\!\mathrm{e}\rangle$) interacting with the electromagnetic field. The atom sits in free space. The Hamiltonian $\hat{H}=\hat{H}_A+\hat{H}_R+\hat{H}_I$ is a sum of three terms: the atom Hamiltonian $\hat{H}_A$, the electromagnetic field Hamiltonian $\hat{H}_R$, and the interaction Hamiltonian $\hat{H}_I$. In the Schrödinger picture these read \cite{CohenQED1}
\begin{subequations} \label{eq:Hamilton}
                \begin{align}
                  \hat{H}_A&=\hbar\omega_{\mathrm{g}}\mid\!\mathrm{g}\rangle\langle \mathrm{g}\!\mid + \hbar\omega_{\mathrm{e}}\mid\!\mathrm{e}\rangle\langle \mathrm{e}\!\mid& \label{eq:HAtom}\\
                  \hat{H}_R&=\sum_{\lambda=\pm}\int\tilde{\mathrm{d}k}\,\hbar c\left|\left|\mathbf{k}\right|\right|\hat{a}_{\left(\lambda\right)}^\dagger\left(\mathbf{k},t=0\right)\hat{a}_{\left(\lambda\right)}\left(\mathbf{k},t=0\right) \label{eq:HField}\\
                  \hat{H}_I&=\frac{e}{m_e}\hat{\mathbf{A}}\left(\hat{\mathbf{x}},0\right)\cdot\hat{\mathbf{p}} \label{eq:HInt}
                \end{align}
              \end{subequations}
where $\omega_{\mathrm{g/e}}$ is the angular frequency of the ground/excited atomic level, $\lambda$ labels the polarisation of the electromagnetic field, $m_e$ is the electron mass. Also, $\hat{\mathbf{x}}$ is the electron position operator and $\hat{\mathbf{p}}$ is the electron linear momentum operator. The field operator $\hat{\mathbf{A}}\left(\mathbf{x},t\right)$ is the vector potential (here we work in the Coulomb gauge), which is expanded over plane waves as
\begin{equation} \label{eq:MaxwellSolHat}
\begin{aligned} [b]
\mathbf{\hat{A}}\left(\mathbf{x},t\right)&=\sqrt{\frac{\hbar}{\epsilon_0c}}\sum_{\lambda=\pm}\int\tilde{\mathrm{d}k}\left[\hat{a}_{\left(\lambda\right)}\left(\mathbf{k},t\right)\bm{\epsilon}_{\left(\lambda\right)}\left(\mathbf{k}\right)\mathrm{e}^{\mathrm{i}\mathbf{k}\cdot\mathbf{x}}\right.\\
&\hspace{80pt}\left.+\hat{a}_{\left(\lambda\right)}^\dagger\left(\mathbf{k},t\right)\bm{\epsilon}_{\left(\lambda\right)}^*\left(\mathbf{k}\right)\mathrm{e}^{-\mathrm{i}\mathbf{k}\cdot\mathbf{x}}\right].
\end{aligned}
\end{equation}
Here
\begin{equation} \label{eq:InDaTilde}
\tilde{\mathrm{d}k}\equiv\frac{\mathrm{d}^4k}{\left(2\pi\right)^4}2\pi\,\delta\left(k_0^2-\mathbf{k}^2\right)\theta\left(k_0\right),
\end{equation}
is the usual volume element on the lightcone \cite{Weinberg1,ItzyksonZuber} (where $\theta$ stands for the Heaviside distribution), which is an invariant under Poincaré transformations. The polarisation vectors $\bm{\epsilon}_{\left(\lambda=\pm1\right)}\left(\mathbf{k}\right)$ are any two mutually orthogonal unit vectors taken in the plane orthogonal to the wave vector $\mathbf{k}$. Finally, we give the commutation relation between the photon ladder operators, which reads
\begin{equation} \label{eq:aaDagger}
\left[\hat{a}_{\left(\varkappa\right)}\left(\mathbf{k}\right),\hat{a}_{\left(\lambda\right)}^\dagger\left(\mathbf{q}\right)\right]=2\left|\left|\mathbf{k}\right|\right|\left(2\pi\right)^3\delta\left(\mathbf{k}-\mathbf{q}\right)\delta_{\varkappa\lambda}.
\end{equation}
We consider spontaneous emission in a vacuum: at $t=0$, the electron is in its excited state, while no photons are present in the field. For such an initial condition, the rotating wave approximation holds \cite{EdouardIsa} and the state of the system at time $t\geq0$ reads
\begin{widetext}
\begin{equation} \label{eq:Ansatz}
          \mid\!\psi\left(t\right)\rangle=c_{\mathrm{e}}\left(t\right)\mathrm{e}^{-\mathrm{i}\omega_{\mathrm{e}}t}\mid\!\mathrm{e},0\rangle+\sum_{\lambda=\pm}\int\tilde{\mathrm{d}k}\,c_{\mathrm{g},\lambda}\left(\mathbf{k},t\right)\mathrm{e}^{-\mathrm{i}\left(\omega_{\mathrm{g}}+ c\left|\left|\mathbf{k}\right|\right|\right) t}\mid\!\mathrm{g},1_{\lambda,\mathbf{k}}\rangle
        \end{equation}
\end{widetext}
where $\mid\!\mathrm{e},0\rangle$ means that the atom is in its excited state and the field contains no photons and $\mid\!\mathrm{g},1_{\lambda,\mathbf{k}}\rangle$ means that the atom is in its ground state and the field contains a photon of wave vector $\mathbf{k}$ and polarisation $\lambda$.\\

Let us turn to the matrix elements of the interaction Hamiltonian in the Hilbert (sub)space spanned by $\mid\!\mathrm{e},0\rangle$ and $\mid\!\mathrm{g},1_{\lambda,\mathbf{k}}\rangle$. These are well-known for the $2\mathrm{p}-1\mathrm{s}$ hydrogen transition. Writing, for this transition, $\mid\!\mathrm{g}\rangle\equiv\mid\!1\mathrm{s}\rangle$ and $\mid\!\mathrm{e}\rangle\equiv\mid\!2\mathrm{p}\,m_2\rangle$, with $m_2$ the magnetic quantum number of the $2\mathrm{p}$ sublevel considered, one has \cite{FacchiPhD}
\begin{widetext}
\begin{equation} \label{eq:MatrixMultipole}
\langle1\mathrm{s},1_{\lambda,\mathbf{k}}\mid\!\hat{H}_I\!\mid\!2\mathrm{p}\,m_2,0\rangle=-\mathrm{i}\sqrt{\frac{\hbar}{\epsilon_0 c}}\frac{\hbar e}{m_e\,a_0}\frac{2^{\frac{9}{2}}}{3^4}\frac{\bm{\epsilon}_{\left(\lambda\right)}^*\left(\mathbf{k}\right)\cdot\bm{\xi}_{m_2}}{\left[1+\left(\frac{2}{3}a_0\left|\left|\mathbf{k}\right|\right|\right)^2\right]^2}
\end{equation}
\end{widetext}
where we introduced the Bohr radius $a_0$. The $\bm{\xi}_{m_{2}}$ are given by
\begin{subequations} \label{eq:Xi}
\begin{align}
\bm{\xi}_0&=\mathbf{e}_z,\\
\bm{\xi}_{\pm1}&=\mp\frac{\mathbf{e}_x\pm\mathrm{i}\mathbf{e}_y}{\sqrt{2}}.
\end{align}
\end{subequations}
In order to derive (\ref{eq:MatrixMultipole}), one must remember the expressions for the electronic wave functions of the $1\mathrm{s}$ and $2\mathrm{p}\,m_2$ sublevels:
\begin{subequations} \label{eq:HLevels}
\begin{align}
\psi_{1\mathrm{s}}\left(\mathbf{x}\right)&=\frac{\exp\left(-\frac{\left|\left|\mathbf{x}\right|\right|}{a_0}\right)}{\sqrt{\pi a_0^3}},\\
\psi_{2\mathrm{p}\,m_2}\left(\mathbf{x}\right)&=\frac{\exp\left(-\frac{\left|\left|\mathbf{x}\right|\right|}{2a_0}\right)}{8\sqrt{\pi a_0^3}}\frac{\sqrt{2}}{a_0}\mathbf{x}\cdot\bm{\xi}_{m_{2}}.
\end{align}
\end{subequations}
Since we are interested in spontaneous emission, we set $c_{\mathrm{e}}\left(t=0\right)=1$ and $\forall\lambda\in\left\{1,2\right\}\forall\mathbf{k}\in\mathbb{R}^3c_{\mathrm{g},\lambda}\left(\mathbf{k},t=0\right)=0$. We want to compute probability amplitudes of emission, namely
\begin{equation} \label{eq:DecaytoSomeMode}
c_{\mathrm{g},\lambda}\left(\mathbf{k},t\right)=\langle\mathrm{g},1_{\lambda,\mathbf{k}}\!\mid\!\hat{U}\left(t\right)\!\mid\!\mathrm{e},0\rangle
\end{equation}
where $\hat{U}\left(t\right)=\exp\left[\left(-\mathrm{i}/\hbar\right)\hat{H}t\right]$ is the evolution operator for the system. From (\ref{eq:Hamilton}) and (\ref{eq:Ansatz}), we get
\begin{widetext}
\begin{subequations} \label{eq:ExactEQ}
\begin{align}
\dot{c}_{\mathrm{e}}\left(t\right)&=-\sum_{\lambda=\pm}\int\frac{\mathrm{d}\mathbf{k}}{\left(2\pi\right)^32\left|\left|\mathbf{k}\right|\right|}G_\lambda^*\left(\mathbf{k}\right)c_{\mathrm{g},\lambda}\left(\mathbf{k},t\right)\mathrm{e}^{-\mathrm{i}\left(c\left|\left|\mathbf{k}\right|\right|-\omega_{\mathrm{e}}+\omega_{\mathrm{g}}\right)t}, \label{eq:ExactEQe}\\
\dot{c}_{\mathrm{g},\lambda}\left(\mathbf{k},t\right)&=-\frac{\mathrm{i}}{\hbar}G_\lambda\left(\mathbf{k}\right)c_{\mathrm{e}}\left(t\right)\mathrm{e}^{\mathrm{i}\left(c\left|\left|\mathbf{k}\right|\right|-\omega_{\mathrm{e}}+\omega_{\mathrm{g}}\right)t} \label{eq:ExactEQg}
\end{align}
\end{subequations}
\end{widetext}
where $G_\lambda\left(\mathbf{k}\right)=\langle1\mathrm{s},1_{\lambda,\mathbf{k}}\mid\!\hat{H}_I\!\mid\!2\mathrm{p}\,m_2,0\rangle$.

\section{The photon wave function} \label{sec:GlauberRule}

The single-photon wave function, hereafter referred to as ``the photon wave function'', is a very useful object for the description of one-photon Fock states of the electromagnetic field either in momentum space \cite{TellementOuf} or, as developed more recently \cite{BBPWF,Sipe,HawtonLorentz,RaymerSmith}, in direct space (and time). In the rest of the paper, we will use Glauber's photon wave function to investigate the spontaneous emission of light during the atomic transition at hand, and focus on causality. This wave function was first introduced by Glauber and Titulaer \cite{Titulaer} in order to characterize correlations of the electromagnetic field in quantum optics.\\

Consider a pure, single-photon state of the electromagnetic field
\begin{equation} \label{eq:RemindOnePhotonState}
\mid\!1,f\left(t\right)\rangle\equiv\sum_\lambda\int\tilde{\mathrm{d}k}\,\bar{f}_\lambda\left(\mathbf{k},t\right)\hat{a}_{\left(\lambda\right)}^\dagger\left(\mathbf{k}\right)\mid\!0\rangle.
\end{equation}
 The photon wave function can be obtained through Glauber's extraction rule which, when states and operators are defined in the Schrödinger picture, reads \cite{HawtonLorentz,BBPWF,RaymerSmith}
\begin{equation} \label{eq:ExtractionRule}
\bm{\psi}_\perp\left(\mathbf{x},t\right)=\langle0\!\mid\hat{\mathbf{E}}_\perp\left(\mathbf{x},0\right)\mid\!1,f\left(t\right)\rangle
\end{equation}
where $\hat{\mathbf{E}}_\perp$ represents the transverse part of the electric field operator defined through
\begin{widetext}
\begin{equation} \label{eq:JustAMomentPos}
\mathbf{\hat{E}}_\perp\left(\mathbf{x},t\right)=\mathrm{i}\sqrt{\frac{\hbar c}{\epsilon_0}}\sum_{\lambda=\pm}\int\tilde{\mathrm{d}k}\left|\left|\mathbf{k}\right|\right|\left[\hat{a}_{\left(\lambda\right)}\left(\mathbf{k},t\right)\bm{\epsilon}_{\left(\lambda\right)}\left(\mathbf{k}\right)\mathrm{e}^{\mathrm{i}\mathbf{k}\cdot\mathbf{x}}-\hat{a}_{\left(\lambda\right)}^\dagger\left(\mathbf{k},t\right)\bm{\epsilon}_{\left(\lambda\right)}^*\left(\mathbf{k}\right)\mathrm{e}^{-\mathrm{i}\mathbf{k}\cdot\mathbf{x}}\right].
\end{equation}
In our problem, the state of the electromagnetic field is not pure, but rather, as seen from (\ref{eq:Ansatz}), entangled with that of the atom. Hence, projecting onto the single photon sector and applying Glauber's extraction rule (\ref{eq:ExtractionRule}), the single-photon wave function reads
\begin{equation} \label{eq:GeneralWaveFunction}
\begin{aligned} [b]
\bm{\psi}_\perp\left(\mathbf{x},t\right)&\equiv\langle\mathrm{g}\left(t\right),0\!\mid\hat{\mathbf{E}}_\perp\left(\mathbf{x},0\right)\mid\psi\left(t\right)\rangle\\
&=\sum_{\lambda=\pm}\int\tilde{\mathrm{d}k}\,\langle0\!\mid\hat{\mathbf{E}}_\perp\left(\mathbf{x},0\right)c_{\mathrm{g},\lambda}\left(\mathbf{k},t\right)\mathrm{e}^{-\mathrm{i}c\left|\left|\mathbf{k}\right|\right|t}\!\mid\!1_{\lambda,\mathbf{k}}\rangle\\
&=\mathrm{i}\sqrt{\frac{\hbar c}{\epsilon_0}}\sum_{\lambda=\pm}\int\tilde{\mathrm{d}k}\left|\left|\mathbf{k}\right|\right|\mathrm{e}^{\mathrm{i}\left(\mathbf{k}\cdot\mathbf{x}-c\left|\left|\mathbf{k}\right|\right|t\right)}c_{\mathrm{g},\lambda}\left(\mathbf{k},t\right)\bm{\epsilon}_{\left(\lambda\right)}\left(\mathbf{k}\right)
\end{aligned}
\end{equation}

\subsection{Formal computation: preliminary steps} \label{subsec:Prelim}

At this point, it comes in handy to notice that (\ref{eq:ExactEQg}) can be formally integrated, yielding
\begin{equation} \label{eq:FormalInt}
c_{\mathrm{g},\lambda}\left(\mathbf{k},t\right)=-\frac{\mathrm{i}}{\hbar}\int_0^t\mathrm{d}t'\,G_\lambda\left(\mathbf{k}\right)c_{\mathrm{e}}\left(t'\right)\mathrm{e}^{\mathrm{i}\left(c\left|\left|\mathbf{k}\right|\right|-\omega_{\mathrm{e}}+\omega_{\mathrm{g}}\right)t'}
\end{equation}
so that the single-photon wave function reads, in the most general case of our problem,
\begin{equation} \label{eq:FromEdouard}
\begin{aligned} [b]
\bm{\psi}_\perp\left(\mathbf{x},t\right)&=\sqrt{\frac{c}{\hbar\epsilon_0}}\sum_{\lambda=\pm}\int\tilde{\mathrm{d}k}\left|\left|\mathbf{k}\right|\right|\mathrm{e}^{\mathrm{i}\left(\mathbf{k}\cdot\mathbf{x}-c\left|\left|\mathbf{k}\right|\right|t\right)}\bm{\epsilon}_{\left(\lambda\right)}\left(\mathbf{k}\right)G_\lambda\left(\mathbf{k}\right)\int_0^t\mathrm{d}t'\,c_{\mathrm{e}}\left(t'\right)\mathrm{e}^{\mathrm{i}\left(c\left|\left|\mathbf{k}\right|\right|-\omega_{\mathrm{e}}+\omega_{\mathrm{g}}\right)t'}\\
&=-\mathrm{i}\frac{2^{\frac{9}{2}}}{3^4}\frac{\hbar e}{\epsilon_0m_ea_0}\sum_{\lambda=\pm}\int\tilde{\mathrm{d}k}\left|\left|\mathbf{k}\right|\right|\mathrm{e}^{\mathrm{i}\left(\mathbf{k}\cdot\mathbf{x}-c\left|\left|\mathbf{k}\right|\right|t\right)}\bm{\epsilon}_{\left(\lambda\right)}\left(\mathbf{k}\right)\frac{\bm{\epsilon}_{\left(\lambda\right)}^*\left(\mathbf{k}\right)\cdot\bm{\xi}_{m_2}}{\left[1+\left(\frac{\left|\left|\mathbf{k}\right|\right|}{k_\mathrm{X}}\right)^2\right]^2}\int_0^t\mathrm{d}t'\,c_{\mathrm{e}}\left(t'\right)\mathrm{e}^{\mathrm{i}\left(c\left|\left|\mathbf{k}\right|\right|-\omega_0\right)t'}\\
\end{aligned}
\end{equation}
\end{widetext}
where we introduced $\omega_0\equiv\omega_{\mathrm{e}}-\omega_{\mathrm{g}}$ and $k_{\mathrm{X}}\equiv3/\left(2a_0\right)$. The unit polarisation vectors obey the closure relation
\begin{equation} \label{eq:PolarSum}
\sum_{\lambda=\pm}\left(\epsilon_{\left(\lambda\right)}^i\right)^*\left(\mathbf{k}\right)\epsilon_{\left(\lambda\right)}^j\left(\mathbf{k}\right)=\delta^{ij}-\frac{k^ik^j}{\mathbf{k}^2}
\end{equation}
so that the wave function now is
\begin{widetext}
\begin{equation} \label{eq:ToMe}
\begin{aligned} [b]
\bm{\psi}_\perp\left(\mathbf{x},t\right)=-\mathrm{i}\frac{2^{\frac{9}{2}}}{3^4}\frac{\hbar e}{\epsilon_0m_ea_0}\int\tilde{\mathrm{d}k}\left|\left|\mathbf{k}\right|\right|\frac{\mathrm{e}^{\mathrm{i}\left(\mathbf{k}\cdot\mathbf{x}-c\left|\left|\mathbf{k}\right|\right|t\right)}}{\left[1+\left(\frac{\left|\left|\mathbf{k}\right|\right|}{k_\mathrm{X}}\right)^2\right]^2}\left(\bm{\xi}_{m_2}-\frac{\bm{\xi}_{m_2}\cdot\mathbf{k}}{\mathbf{k}^2}\,\mathbf{k}\right)\int_0^t\mathrm{d}t'\,c_{\mathrm{e}}\left(t'\right)\mathrm{e}^{\mathrm{i}\left(c\left|\left|\mathbf{k}\right|\right|-\omega_0\right)t'}.
\end{aligned}
\end{equation}
Choosing a coordinate system for which $\mathbf{x}$ points along the third axis $\mathbf{e}_z$, we can compute the angular integrals:
\begin{align*}
\mathbf{F}\left(k,\left|\left|\mathbf{x}\right|\right|\right)&\equiv\int_0^\pi\mathrm{d}\theta\sin\theta\int_0^{2\pi}\mathrm{d}\varphi\left[\bm{\xi}_{m_2}-\left(\bm{\xi}_{m_2}\cdot\frac{\mathbf{k}}{\left|\left|\mathbf{k}\right|\right|}\right)\frac{\mathbf{k}}{\left|\left|\mathbf{k}\right|\right|}\right]\mathrm{e}^{\mathrm{i}k\left|\left|\mathbf{x}\right|\right|\cos\theta}\\
&=\int_0^\pi\mathrm{d}\theta\sin\theta\,\mathrm{e}^{\mathrm{i}k\left|\left|\mathbf{x}\right|\right|\cos\theta}\int_0^{2\pi}\mathrm{d}\varphi\left\{\left[\begin{array}{c}\xi_{m_2}^{\left(x\right)}\\\xi_{m_2}^{\left(y\right)}\\\xi_{m_2}^{\left(z\right)}\end{array}\right]-\left[\begin{array}{c}\sin\theta\cos\varphi\\\sin\theta\sin\varphi\\\cos\theta\end{array}\right]\left(\xi_{m_2}^{\left(x\right)}\sin\theta\cos\varphi+\xi_{m_2}^{\left(y\right)}\sin\theta\sin\varphi+\xi_{m_2}^{\left(z\right)}\cos\theta\right)\vphantom{\left[\begin{array}{c}\xi_{m_2}^{\left(1\right)}\\\xi_{m_2}^{\left(2\right)}\\\xi_{m_2}^{\left(z\right)}\end{array}\right]}\right\}\\
&=2\pi\int_0^\pi\mathrm{d}\theta\sin\theta\,\mathrm{e}^{\mathrm{i}k\left|\left|\mathbf{x}\right|\right|\cos\theta}\left[\begin{array}{c}\xi_{m_2}^{\left(x\right)}\left(1-\frac{1}{2}\sin^2\theta\right)\\\\\xi_{m_2}^{\left(y\right)}\left(1-\frac{1}{2}\sin^2\theta\right)\\\\\xi_{m_2}^{\left(z\right)}\sin^2\theta\end{array}\right]\\
&\equiv2\pi\,\mathbf{I}\left(k,\left|\left|\mathbf{x}\right|\right|\right).
\end{align*}
The integrals over $\theta$ give
\begin{align} \label{eq:DipoleGreen}
I^{\left(x,y\right)}\left(k,\left|\left|\mathbf{x}\right|\right|\right)&=\mathrm{i}\frac{\xi_{m_2}^{\left(x,y\right)}}{k\left|\left|\mathbf{x}\right|\right|}&\left[\vphantom{\frac{1}{\left(k\left|\left|\mathbf{x}\right|\right|\right)^2}}\right.&\left(\mathrm{e}^{-\mathrm{i}k\left|\left|\mathbf{x}\right|\right|}-\mathrm{e}^{\mathrm{i}k\left|\left|\mathbf{x}\right|\right|}\right)&-\frac{\mathrm{i}}{k\left|\left|\mathbf{x}\right|\right|}\left(\mathrm{e}^{-\mathrm{i}k\left|\left|\mathbf{x}\right|\right|}+\mathrm{e}^{\mathrm{i}k\left|\left|\mathbf{x}\right|\right|}\right)&\left.-\frac{1}{\left(k\left|\left|\mathbf{x}\right|\right|\right)^2}\left(\mathrm{e}^{-\mathrm{i}k\left|\left|\mathbf{x}\right|\right|}-\mathrm{e}^{\mathrm{i}k\left|\left|\mathbf{x}\right|\right|}\right)\right],\\
I^{\left(z\right)}\left(k,\left|\left|\mathbf{x}\right|\right|\right)&=-2\mathrm{i}\frac{\xi_{m_2}^{\left(z\right)}}{k\left|\left|\mathbf{x}\right|\right|}&\left[\vphantom{\frac{1}{\left(k\left|\left|\mathbf{x}\right|\right|\right)^2}}\right.&&-\frac{\mathrm{i}}{k\left|\left|\mathbf{x}\right|\right|}\left(\mathrm{e}^{-\mathrm{i}k\left|\left|\mathbf{x}\right|\right|}+\mathrm{e}^{\mathrm{i}k\left|\left|\mathbf{x}\right|\right|}\right)&\left.-\frac{1}{\left(k\left|\left|\mathbf{x}\right|\right|\right)^2}\left(\mathrm{e}^{-\mathrm{i}k\left|\left|\mathbf{x}\right|\right|}-\mathrm{e}^{\mathrm{i}k\left|\left|\mathbf{x}\right|\right|}\right)\right].
\end{align}
As we can see, the photon wave function contains contributions proportional to $1/\left|\left|\mathbf{x}\right|\right|$, $1/\left|\left|\mathbf{x}\right|\right|^2$ and $1/\left|\left|\mathbf{x}\right|\right|^3$, which are respectively known as the far-field, mid-field and near-field contributions. We then have
\begin{equation} \label{eq:PsiBeta}
\bm{\psi}_\perp\left(\mathbf{x},t\right)=-\mathrm{i}\frac{2^{\frac{7}{2}}}{3^4}\frac{\hbar e}{\epsilon_0m_ea_0}\int_0^{+\infty}\frac{\mathrm{d}k}{\left(2\pi\right)^2}k^2\frac{\mathrm{e}^{-\mathrm{i}ckt}}{\left[1+\left(\frac{k}{k_\mathrm{X}}\right)^2\right]^2}\mathbf{I}\left(k,\left|\left|\mathbf{x}\right|\right|\right)\int_0^t\mathrm{d}t'\,c_{\mathrm{e}}\left(t'\right)\mathrm{e}^{\mathrm{i}\left(ck-\omega_0\right)t'}.
\end{equation}
\end{widetext}
In most of what follows we will use the Wigner-Weisskopf approximation of exponential decay. We shall then have
\begin{equation}
c_{\mathrm{e}}\left(t\right)=\mathrm{e}^{-\mathrm{i}\omega_{\mathrm{LS}}t}\,\mathrm{e}^{-\frac{1}{2}\Gamma t}
\end{equation}
where $\omega_{\mathrm{LS}}$ is the partial Lamb shift \cite{FacchiPascazio} of the excited $2\mathrm{p}$ level due to the $1\mathrm{s}$ level and $\Gamma$ is the decay rate. This yields
{\allowdisplaybreaks
\begin{equation} \label{eq:WWInt}
\begin{aligned} [b]
\int_0^t\mathrm{d}t'\,c_{\mathrm{e}}\left(t'\right)\mathrm{e}^{\mathrm{i}\left(ck-\omega_0\right)t'}&=\int_0^t\mathrm{d}t'\,\mathrm{e}^{\mathrm{i}\left(ck-\left(\omega_0+\omega_{\mathrm{LS}}\right)\right)t'}\,\mathrm{e}^{-\frac{1}{2}\Gamma t'}\\
&=\left[\frac{\mathrm{e}^{\mathrm{i}\left(ck-\left(\omega_0+\omega_{\mathrm{LS}}\right)\right)t'}\,\mathrm{e}^{-\frac{1}{2}\Gamma t'}}{\mathrm{i}\left(ck-\left(\omega_0+\omega_{\mathrm{LS}}\right)\right)-\frac{1}{2}\Gamma}\right]_0^t\\
&=\mathrm{i}\,\frac{1-\mathrm{e}^{\mathrm{i}\left(ck-\left(\omega_0+\omega_{\mathrm{LS}}\right)\right)t}\,\mathrm{e}^{-\frac{1}{2}\Gamma t}}{ck-\left(\omega_0+\omega_{\mathrm{LS}}\right)+\frac{\mathrm{i}}{2}\Gamma}.
\end{aligned}
\end{equation}}
We introduce the space-saving notation,
\begin{equation}
\Omega_0\equiv\omega_0+\omega_{\mathrm{LS}}-\frac{\mathrm{i}}{2}\Gamma,
\end{equation}
We can then write the contributions to the far-field, mid-field, and near-field to this photon wave function:
\begin{widetext}
\begin{subequations} \label{eq:GiveMeaMoment}
\begin{align} \label{eq:NobodyCares}
\psi_{\perp\left(\mathrm{far}\right)}^{\left(x,y\right)}&=\mathrm{i}\frac{2^{\frac{7}{2}}}{3^4}\frac{\hbar e}{\epsilon_0m_ea_0}\frac{\xi_{m_2}^{\left(x,y\right)}}{\left|\left|\mathbf{x}\right|\right|}\int_0^{+\infty}\frac{\mathrm{d}k}{\left(2\pi\right)^2}k\frac{\mathrm{e}^{-\mathrm{i}\Omega_0t}}{\left[1+\left(\frac{k}{k_\mathrm{X}}\right)^2\right]^2}\left(\mathrm{e}^{-\mathrm{i}k\left|\left|\mathbf{x}\right|\right|}-\mathrm{e}^{\mathrm{i}k\left|\left|\mathbf{x}\right|\right|}\right)\frac{1-\mathrm{e}^{-\mathrm{i}\left(ck-\Omega_0\right)t}}{ck-\Omega_0},\\
\psi_{\perp\left(\mathrm{mid}\right)}^{\left(x,y\right)}&=\frac{2^{\frac{7}{2}}}{3^4}\frac{\hbar e}{\epsilon_0m_ea_0}\frac{\xi_{m_2}^{\left(x,y\right)}}{\left|\left|\mathbf{x}\right|\right|^2}\int_0^{+\infty}\frac{\mathrm{d}k}{\left(2\pi\right)^2}\frac{\mathrm{e}^{-\mathrm{i}\Omega_0t}}{\left[1+\left(\frac{k}{k_\mathrm{X}}\right)^2\right]^2}\left(\mathrm{e}^{-\mathrm{i}k\left|\left|\mathbf{x}\right|\right|}+\mathrm{e}^{\mathrm{i}k\left|\left|\mathbf{x}\right|\right|}\right)\frac{1-\mathrm{e}^{-\mathrm{i}\left(ck-\Omega_0\right)t}}{ck-\Omega_0},\\
\psi_{\perp\left(\mathrm{near}\right)}^{\left(x,y\right)}&=-\mathrm{i}\frac{2^{\frac{7}{2}}}{3^4}\frac{\hbar e}{\epsilon_0m_ea_0}\frac{\xi_{m_2}^{\left(x,y\right)}}{\left|\left|\mathbf{x}\right|\right|^3}\int_0^{+\infty}\frac{\mathrm{d}k}{\left(2\pi\right)^2}\frac{1}{k}\frac{\mathrm{e}^{-\mathrm{i}\Omega_0t}}{\left[1+\left(\frac{k}{k_\mathrm{X}}\right)^2\right]^2}\left(\mathrm{e}^{-\mathrm{i}k\left|\left|\mathbf{x}\right|\right|}-\mathrm{e}^{\mathrm{i}k\left|\left|\mathbf{x}\right|\right|}\right)\frac{1-\mathrm{e}^{-\mathrm{i}\left(ck-\Omega_0\right)t}}{ck-\Omega_0},\\
\psi_{\perp\left(\mathrm{far}\right)}^{\left(z\right)}&=0,\\
\psi_{\perp\left(\mathrm{mid}\right)}^{\left(z\right)}&=-\frac{2^{\frac{9}{2}}}{3^4}\frac{\hbar e}{\epsilon_0m_ea_0}\frac{\xi_{m_2}^{\left(z\right)}}{\left|\left|\mathbf{x}\right|\right|^2}\int_0^{+\infty}\frac{\mathrm{d}k}{\left(2\pi\right)^2}\frac{\mathrm{e}^{-\mathrm{i}\Omega_0t}}{\left[1+\left(\frac{k}{k_\mathrm{X}}\right)^2\right]^2}\left(\mathrm{e}^{-\mathrm{i}k\left|\left|\mathbf{x}\right|\right|}+\mathrm{e}^{\mathrm{i}k\left|\left|\mathbf{x}\right|\right|}\right)\frac{1-\mathrm{e}^{-\mathrm{i}\left(ck-\Omega_0\right)t}}{ck-\Omega_0},\\
\psi_{\perp\left(\mathrm{near}\right)}^{\left(z\right)}&=\mathrm{i}\frac{2^{\frac{9}{2}}}{3^4}\frac{\hbar e}{\epsilon_0m_ea_0}\frac{\xi_{m_2}^{\left(z\right)}}{\left|\left|\mathbf{x}\right|\right|^3}\int_0^{+\infty}\frac{\mathrm{d}k}{\left(2\pi\right)^2}\frac{1}{k}\frac{\mathrm{e}^{-\mathrm{i}\Omega_0t}}{\left[1+\left(\frac{k}{k_\mathrm{X}}\right)^2\right]^2}\left(\mathrm{e}^{-\mathrm{i}k\left|\left|\mathbf{x}\right|\right|}-\mathrm{e}^{\mathrm{i}k\left|\left|\mathbf{x}\right|\right|}\right)\frac{1-\mathrm{e}^{-\mathrm{i}\left(ck-\Omega_0\right)t}}{ck-\Omega_0}.
\end{align}
\end{subequations}

\subsection{General method} \label{subsec:NotationsGalore}

We write
\begin{equation} \label{eq:ThisIsK}
\begin{aligned} [b]
H_n^{\left(\pm\right)}\left(\left|\left|\mathbf{x}\right|\right|,t\right)&\equiv\int_0^{+\infty}\mathrm{d}k\,\frac{k^{2-n}}{\left[1+\left(\frac{k}{k_{\mathrm{X}}}\right)^2\right]^2}\mathrm{e}^{\pm\mathrm{i}k\left|\left|\mathbf{x}\right|\right|}\,\frac{1-\mathrm{e}^{-\mathrm{i}\left(ck-\omega_0\right)t}}{ck-\omega_0}\\
&\equiv\int_0^{+\infty}\mathrm{d}k\,f_n\left(k\right)\mathrm{e}^{\pm\mathrm{i}k\left|\left|\mathbf{x}\right|\right|}.
\end{aligned}
\end{equation}
It is a general result of distribution theory \cite{Appel} that
\begin{equation} \label{eq:ConvWithDeltavpFuture}
\begin{aligned} [b]
H_n^{\left(\pm\right)}\left(\left|\left|\mathbf{x}\right|\right|,t\right)&=\frac{1}{2}\left[\left(\delta\left(\cdot\right)-\frac{\mathrm{i}}{\pi}\mathrm{vp}\,\frac{1}{\cdot}\right)*\bar{f}_n\left(\cdot,t\right)\right]\left(\mp\left|\left|\mathbf{x}\right|\right|\right)\\
&\equiv\frac{1}{2}\bar{f}_n\left(\mp\left|\left|\mathbf{x}\right|\right|,t\right)-\frac{\mathrm{i}}{2\pi}\lim_{\epsilon\to0^+}\left[\int_{-\infty}^{-\epsilon}+\int_\epsilon^{+\infty}\right]\frac{\mathrm{d}\sigma}{\mp\left|\left|\mathbf{x}\right|\right|-\sigma}\bar{f}_n\left(\sigma,t\right)
\end{aligned}
\end{equation}
\end{widetext}
with the Fourier transform
\begin{equation} \label{eq:SortStart}
\bar{f}_n\left(\left|\left|\mathbf{x}\right|\right|,t\right)=\int_{-\infty}^{+\infty}\mathrm{d}k\,f_n\left(k,t\right)\mathrm{e}^{-\mathrm{i}kx}.
\end{equation}
In what follows we will endeavour to compute the $H_n^{\left(\pm\right)}$ integrals, first by making use of several aproximations in order to obtain a causal result, and thereafter progressively waiving these approximations.

\section{Causality in the standard treatment} \label{sec:CausalWW}

Here we use the following standard \cite{FermiCausal,FoxMulder} approximations used to established the causality of the wave function of the emitted photon: the Wigner-Weisskopf exponential decay, the usual $\hat{\mathbf{E}}\cdot\hat{\mathbf{x}}$ coupling between the field and the atom instead of the minimal $\hat{\mathbf{A}}\cdot\hat{\mathbf{p}}$ coupling, the dipole approximation and the approximation which consists, as we will see, in extending the range of electromagnetic frequencies to the negative real semi-axis.\\

We switch (only in the present section) from the minimal $\hat{\mathbf{A}}\cdot\hat{\mathbf{p}}$ to the usual $\hat{\mathbf{E}}\cdot\hat{\mathbf{x}}$ coupling. In the dipole approximation, which, for the $\hat{\mathbf{A}}\cdot\hat{\mathbf{p}}$ coupling, consists \cite{FredZ} in forgetting about the $\left[1+\left(k/k_{\mathrm{X}}\right)^2\right]^2$ denominator in (\ref{eq:PsiBeta}), this substitution results \cite{EdouardIsa,CohenQED1} in the multiplication of the interaction matrix element by $ck/\omega_0$. Plugging this in (\ref{eq:PsiBeta}), we have
\begin{multline} \label{eq:PreTrick}
\bm{\psi}_\perp\left(\mathbf{x},t\right)=-\frac{2^{\frac{7}{2}}}{3^4}\frac{\hbar ec}{\epsilon_0m_ea_0\omega_0}\int_0^{+\infty}\frac{\mathrm{d}k}{\left(2\pi\right)^2}k^3\mathrm{e}^{-\mathrm{i}\Omega_0t}\\
\mathbf{I}\left(k,\left|\left|\mathbf{x}\right|\right|\right)\frac{1-\mathrm{e}^{-\mathrm{i}\left(ck-\Omega_0\right)t}}{ck-\Omega_0}.
\end{multline}
The usual trick \cite{FermiCausal,Shirokov,FoxMulder} is then to extend the range of integration from the positive real semi-axis to the whole real axis, and we find ourselves computing integrals of the type
\begin{equation} \label{eq:IntheDark}
H_{n\left(\mathrm{std}\right)}^{\left(\pm\right)}\left(\left|\left|\mathbf{x}\right|\right|,t\right)\equiv\int_{-\infty}^{+\infty}\mathrm{d}k\,k^{3-n}\mathrm{e}^{\pm\mathrm{i}k\left|\left|\mathbf{x}\right|\right|}\,\frac{1-\mathrm{e}^{-\mathrm{i}\left(ck-\Omega_0\right)t}}{ck-\Omega_0},
\end{equation}
with $n\in\left\{1,2,3\right\}$, and where the label $+$($-$) is assigned to outgoing (ingoing) radial waves. Here the $\mathrm{std}$ subscript stands for ``standard'' as we follow the lines of the standard treatment \cite{FermiCausal,FoxMulder} of the problem. 
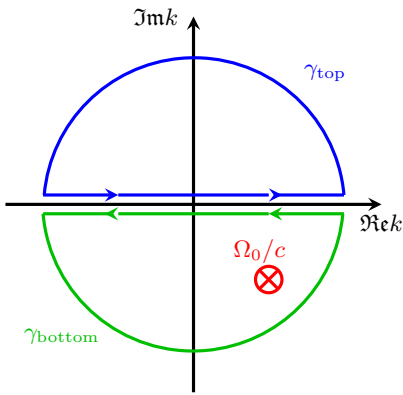
\begin{figure}[t]
\begin{center}
\begin{tikzpicture}[very thick, >=stealth]
\draw[->] (-2.5,0) -- (2.5,0);
\draw[->] (0,-2.5) -- (0,2.5);
\draw (2.5,-.25) node {\small $\mathfrak{Re}k$};
\draw (-.5,2.5) node {\small $\mathfrak{Im}k$};
\draw[red] (.875,-1.125) -- (1.125,-.875);
\draw[red] (.875,-.875) -- (1.125,-1.125);
\draw[red] (1,-1) circle (5pt);
\draw[red] (.875,-.625) node {\small $\Omega_0/c$};
\draw[blue, ->] (-2,.125) -- (-1,.125);
\draw[blue] (-1,.125) -- (1,.125);
\draw[blue, >-] (1,.125) -- (2,.125);
\draw[blue] (2,.125) arc (5:175:2);
\draw[blue] (1.75,1.75) node {\small $\gamma_{\mathrm{top}}$};
\draw[green!75!black, ->] (2,-.125) -- (1,-.125);
\draw[green!75!black] (1,-.125) -- (-1,-.125);
\draw[green!75!black, >-] (-1,-.125) -- (-2,-.125);
\draw[green!75!black] (-2,-.125) arc (185:355:2);
\draw[green!75!black] (-1.75,-1.75) node {\small $\gamma_{\mathrm{bottom}}$};
\end{tikzpicture}
\end{center}
\vspace{-10pt}
\caption{Jordan loops in the complex $k$-plane used to compute the integrals (\ref{eq:IntheDark}). The (isolated) simple pole $\Omega_0/c$ of the integrands is represented by a red circled cross. \label{fig:JLoop}}
\end{figure}
For this we use Cauchy's residue theorem (see Fig.~\ref{fig:JLoop}). Taking into account the fact that $t$ and $\left|\left|\mathbf{x}\right|\right|$ are positive quantities, we find (for more details, see the similar treatment of sect.~\ref{subsec:LongPole})
\begin{equation} \label{eq:FermiIntegrals}
\begin{aligned} [b]
H_{n\left(\mathrm{std}\right)}^{\left(+\right)}\left(\left|\left|\mathbf{x}\right|\right|,t\right)&=2\frac{\mathrm{i}\pi}{c}\,\theta\left(ct-\left|\left|\mathbf{x}\right|\right|\right)\left(\frac{\Omega_0}{c}\right)^{3-n}\mathrm{e}^{\frac{\mathrm{i}}{c}\Omega_0\left|\left|\mathbf{x}\right|\right|},\\
H_{n\left(\mathrm{std}\right)}^{\left(-\right)}\left(\left|\left|\mathbf{x}\right|\right|,t\right)&=0.
\end{aligned}
\end{equation}
This means that contributions from ingoing waves are zero, as found for instance in \cite{FoxMulder}. As made clear by the Heaviside step, this result is explicitly causal. We finally have
\begin{widetext}
\begin{equation} \label{eq:HadtoAdd}
\bm{\psi}_\perp\left(\mathbf{x},t\right)=\frac{2^{\frac{5}{2}}}{3^4\pi}\frac{\hbar e}{\epsilon_0m_ea_0\omega_0}\,\theta\left(ct-\left|\left|\mathbf{x}\right|\right|\right)\left(\frac{\Omega_0}{c}\right)^3\left[\begin{array}{crcccl}\xi^{\left(x,y\right)}&\left(\vphantom{\frac{1}{\left(\frac{\Omega_0}{c}\left|\left|\mathbf{x}\right|\right|\right)^3}}\right.&\frac{1}{\frac{\Omega_0}{c}\left|\left|\mathbf{x}\right|\right|}&+\frac{\mathrm{i}}{\left(\frac{\Omega_0}{c}\left|\left|\mathbf{x}\right|\right|\right)^2}&-\frac{1}{\left(\frac{\Omega_0}{c}\left|\left|\mathbf{x}\right|\right|\right)^3}&\left.\vphantom{\frac{1}{\left(\frac{\Omega_0}{c}\left|\left|\mathbf{x}\right|\right|\right)^3}}\right)\\-2\xi^{\left(z\right)}&\left(\vphantom{\frac{1}{\left(\frac{\Omega_0}{c}\left|\left|\mathbf{x}\right|\right|\right)^3}}\right.&&+\frac{\mathrm{i}}{\left(\frac{\Omega_0}{c}\left|\left|\mathbf{x}\right|\right|\right)^2}&-\frac{1}{\left(\frac{\Omega_0}{c}\left|\left|\mathbf{x}\right|\right|\right)^3}&\left.\vphantom{\frac{1}{\left(\frac{\Omega_0}{c}\left|\left|\mathbf{x}\right|\right|\right)^3}}\right)\end{array}\right]\mathrm{e}^{\mathrm{i}\frac{\Omega_0}{c}\left(\left|\left|\mathbf{x}\right|\right|-ct\right)}.
\end{equation}
\end{widetext}

\section{Minimal coupling in the dipole approximation} \label{sec:PreStorm}

\subsection{Intrinsic noncausality in the near-field} \label{subsec:LongPole}

We now return to what we regard as the more correct coupling: the minimal $\hat{\mathbf{A}}\cdot\hat{\mathbf{p}}$ coupling. We still work in the dipole approximation and in the Wigner-Weisskopf approximation. We shall first extend the range of integration to the negative real semi-axis as was done in the previous sect.~\ref{sec:CausalWW}. This yields
\begin{multline} \label{eq:Patrick}
\bm{\psi}_\perp\left(\mathbf{x},t\right)=-\frac{2^{\frac{7}{2}}}{3^4}\frac{\hbar e}{\epsilon_0m_ea_0}\int_{-\infty}^{+\infty}\frac{\mathrm{d}k}{\left(2\pi\right)^2}k^2\mathrm{e}^{-\mathrm{i}\Omega_0t}\\
\mathbf{I}\left(k,\left|\left|\mathbf{x}\right|\right|\right)\frac{1-\mathrm{e}^{-\mathrm{i}\left(ck-\Omega_0\right)t}}{ck-\Omega_0}.
\end{multline}
We find ourselves computing the following Fourier transforms:
\begin{equation} \label{eq:Bob}
\begin{aligned} [b]
\bar{f}_{n\left(\mathrm{dip}\right)}\left(\mp\left|\left|\mathbf{x}\right|\right|,t\right)&\equiv\int_{-\infty}^{+\infty}\mathrm{d}k\,k^{2-n}\mathrm{e}^{\pm\mathrm{i}k\left|\left|\mathbf{x}\right|\right|}\,\frac{1-\mathrm{e}^{-\mathrm{i}\left(ck-\Omega_0\right)t}}{ck-\Omega_0}\\
&\equiv\int_{-\infty}^{+\infty}\mathrm{d}k\,f_{n\left(\mathrm{dip}\right)}\left(k\right)\mathrm{e}^{\pm\mathrm{i}k\left|\left|\mathbf{x}\right|\right|}.
\end{aligned}
\end{equation}
Here the $\mathrm{dip}$ subscript stands for ``dipole'' as we use the dipole-approximated minimal $\hat{\mathbf{A}}\cdot\hat{\mathbf{p}}$ coupling.
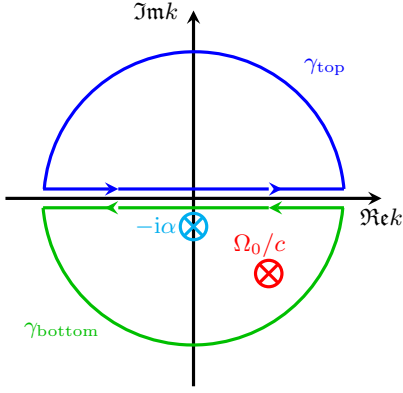
\begin{figure}[t]
\begin{center}
\begin{tikzpicture}[very thick, >=stealth]
\draw[->] (-2.5,0) -- (2.5,0);
\draw[->] (0,-2.5) -- (0,2.5);
\draw (2.5,-.25) node {\small $\mathfrak{Re}k$};
\draw (-.5,2.5) node {\small $\mathfrak{Im}k$};
\draw[red] (.875,-1.125) -- (1.125,-.875);
\draw[red] (.875,-.875) -- (1.125,-1.125);
\draw[red] (1,-1) circle (5pt);
\draw[red] (.875,-.625) node {\small $\Omega_0/c$};
\draw[cyan] (-.125,-.5) -- (.125,-.25);
\draw[cyan] (-.125,-.25) -- (.125,-.5);
\draw[cyan] (0,-.375) circle (5pt);
\draw[cyan] (-.5,-.375) node {\small $-\mathrm{i}\alpha$};
\draw[blue, ->] (-2,.125) -- (-1,.125);
\draw[blue] (-1,.125) -- (1,.125);
\draw[blue, >-] (1,.125) -- (2,.125);
\draw[blue] (2,.125) arc (5:175:2);
\draw[blue] (1.75,1.75) node {\small $\gamma_{\mathrm{top}}$};
\draw[green!75!black, ->] (2,-.125) -- (1,-.125);
\draw[green!75!black] (1,-.125) -- (-1,-.125);
\draw[green!75!black, >-] (-1,-.125) -- (-2,-.125);
\draw[green!75!black] (-2,-.125) arc (185:355:2);
\draw[green!75!black] (-1.75,-1.75) node {\small $\gamma_{\mathrm{bottom}}$};
\end{tikzpicture}
\end{center}
\vspace{-10pt}
\caption{Jordan loops in the complex $k$-plane used to compute the integrals (\ref{eq:Bob}). The (isolated) simple poles $\Omega_0/c$ and $-\mathrm{i}\alpha$ of the integrands are represented by a red and a cyan circled cross, respectively. \label{fig:Camille}}
\end{figure}
We use Cauchy's residue theorem (see Fig.~\ref{fig:Camille}). Taking into account the fact that $t$ and $\left|\left|\mathbf{x}\right|\right|$ are positive quantities, we find, for $n\in\left\{1,2\right\}$
\begin{widetext}
\begin{subequations} \label{eq:UXWhatever}
\begin{equation} \label{eq:UArb}
\frac{\bar{f}_{n\left(\mathrm{dip}\right)}\left(u,t\right)}{2\mathrm{i}\pi}=\left(\frac{\Omega_0}{c}\right)^{2-n}\mathrm{e}^{-\mathrm{i}\frac{\Omega_0}{c}u}\left[-\theta\left(u\right)+\theta\left(u+ct\right)\right]
\end{equation}
whence
\begin{equation} \label{eq:FermiIntegral}
\bar{f}_{n\left(\mathrm{dip}\right)}\left(\left|\left|\mathbf{x}\right|\right|,t\right)=0
\end{equation}
and
\begin{equation} \label{eq:FermiIntegralMinus}
\bar{f}_{n\left(\mathrm{dip}\right)}\left(-\left|\left|\mathbf{x}\right|\right|,t\right)=2\mathrm{i}\pi\left(\frac{\Omega_0}{c}\right)^{2-n}\mathrm{e}^{\mathrm{i}\frac{\Omega_0}{c}\left|\left|\mathbf{x}\right|\right|}\theta\left(-\left|\left|\mathbf{x}\right|\right|+ct\right),
\end{equation}
from which we conclude
\begin{equation} \label{eq:PlusMinusMinusY}
\bar{f}_{n\left(\mathrm{dip}\right)}\left(-\left|\left|\mathbf{x}\right|\right|,t\right)\mp\bar{f}_{n\left(\mathrm{dip}\right)}\left(\left|\left|\mathbf{x}\right|\right|,t\right)=2\mathrm{i}\pi\left(\frac{\Omega_0}{c}\right)^{2-n}\mathrm{e}^{\mathrm{i}\frac{\Omega_0}{c}\left|\left|\mathbf{x}\right|\right|}\theta\left(-\left|\left|\mathbf{x}\right|\right|+ct\right)
\end{equation}
\end{subequations}
\end{widetext}
which is quite transparently causal (notice from (\ref{eq:FermiIntegral}) that the contributions from incoming waves are identically zero here, as they were in the standard treatment of sect.~\ref{sec:CausalWW}). Now, notice that for $n=3$, the integrand in (\ref{eq:Bob}) has a (simple) pole at $k=0$ (see Fig.~\ref{fig:Camille}). Since we integrate over the real axis, this could be a serious problem, because it means that $J_3^{\left(\pm\right)}$ are divergent integrals. We are saved from dealing with such divergences by noticing that, according to (\ref{eq:DipoleGreen}), we will only be interested in computing the difference $\left(J_3^{\left(+\right)}-J_3^{\left(-\right)}\right)\left(\left|\left|\mathbf{x}\right|\right|,t\right)$, which is an integral over a function which, according to (\ref{eq:Bob}), has only an artificial singularity at $k=0$ \footnote{Let us note that, for $n\in\left\{1,2\right\}$, the integrand in (\ref{eq:Bob}) actually is an entire function of $k$, since the singularity at $\Omega_0/c$ is also artificial. To compute the integral, though, we split the integrand into two meromorphic functions in order to be able to use the Jordan lemma.}. Accordingly, we can shift this singularity away from the real axis $0\rightarrow-\mathrm{i}\alpha$ to compute the integral (see Fig.~\ref{fig:Camille}), before taking the limit $\alpha\rightarrow0$ at the end. This amounts to performing the substitution $k^{2-n}\rightarrow\left(k+\mathrm{i}\alpha\right)^{2-n}$ and yields
\begin{widetext}
\begin{equation} \label{eq:ThreePlusMinusMinus}
\frac{\bar{f}_{3\left(\mathrm{dip}\right)}\left(-\left|\left|\mathbf{x}\right|\right|,t\right)-\bar{f}_{3\left(\mathrm{dip}\right)}\left(\left|\left|\mathbf{x}\right|\right|,t\right)}{2\mathrm{i}\pi}=\frac{c}{\Omega_0}\left[\theta\left(ct-\left|\left|\mathbf{x}\right|\right|\right)\left(\mathrm{e}^{\frac{\mathrm{i}}{c}\Omega_0\left|\left|\mathbf{x}\right|\right|}-\mathrm{e}^{\mathrm{i}\Omega_0t}\right)+\left(1-\mathrm{e}^{\mathrm{i}\Omega_0t}\right)\vphantom{\theta\left(ct-\left|\left|\mathbf{x}\right|\right|\right)\left(\mathrm{e}^{\frac{\mathrm{i}}{c}\Omega_0\left|\left|\mathbf{x}\right|\right|}-\mathrm{e}^{\mathrm{i}\Omega_0t}\right)}\right]
\end{equation}
\end{widetext}
Comparison with (\ref{eq:PlusMinusMinusY}) shows that the result for $n=3$ features not only an extra causal contribution, but also a completely noncausal term. This is a feature of the slow dependence of the $\hat{\mathbf{A}}\cdot\hat{\mathbf{p}}$ atom-field coupling on the norm of the electromagnetic wave vector $\mathbf{k}$. As far as we know, similar calculations \cite{FermiCausal,FoxMulder,PassanteVirtual,Sipe} of the outgoing field have mostly been carried out in the Power-Zineau-Woolley picture \cite{CohenQED1} of quantum electrodynamics where the interaction Hamiltonian is of the usual $\hat{\mathbf{E}}\cdot\hat{\mathbf{x}}$ form. In this case no pole is present at $k=0$ (see Figs.~\ref{fig:JLoop} and \ref{fig:Camille}) and one retrieves a causal result, as was done in sect.~\ref{sec:CausalWW}.

\subsection{Hegerfeldt theorem for the minimal coupling in the dipole approximation} \label{subsec:AdotPDipHeg}

We now waive the approximation which consists in extending the range of integration to the negative real semi-axis as was done in the previous sect.~\ref{sec:CausalWW}. This yields
\begin{multline} \label{eq:Patrick}
\bm{\psi}_\perp\left(\mathbf{x},t\right)=-\frac{2^{\frac{7}{2}}}{3^4}\frac{\hbar e}{\epsilon_0m_ea_0}\int_0^{+\infty}\frac{\mathrm{d}k}{\left(2\pi\right)^2}k^2\mathrm{e}^{-\mathrm{i}\Omega_0t}\\
\mathbf{I}\left(k,\left|\left|\mathbf{x}\right|\right|\right)\frac{1-\mathrm{e}^{-\mathrm{i}\left(ck-\Omega_0\right)t}}{ck-\Omega_0}
\end{multline}
and we find ourselves computing integrals of the type
\begin{equation} \label{eq:BobAgain}
\begin{aligned} [b]
H_{n\left(\mathrm{dip}\right)}^{\left(\pm\right)}\left(\left|\left|\mathbf{x}\right|\right|,t\right)&\equiv\int_0^{+\infty}\mathrm{d}k\,k^{2-n}\mathrm{e}^{\pm\mathrm{i}k\left|\left|\mathbf{x}\right|\right|}\,\frac{1-\mathrm{e}^{-\mathrm{i}\left(ck-\Omega_0\right)t}}{ck-\Omega_0}\\
&\equiv\int_{-\infty}^{+\infty}\mathrm{d}k\,\theta\left(k\right)f_{n\left(\mathrm{dip}\right)}\left(k\right)\mathrm{e}^{\pm\mathrm{i}k\left|\left|\mathbf{x}\right|\right|}
\end{aligned}
\end{equation}
with $n\in\left\{1,2,3\right\}$. With (\ref{eq:ConvWithDeltavpFuture}) in mind we can use the Fourier transform $\bar{f}_{n\left(\mathrm{dip}\right)}$ of $f_{n\left(\mathrm{dip}\right)}$ computed in sect.~\ref{subsec:LongPole}. Since we did not extend the range of integration to the whole real axis, we will not retrieve a causal propagation, as first noted by Shirokov \cite{Shirokov}. Note that this is true regardless of the choice for the coupling: Fermi's proof of causality used the usual $\hat{\mathbf{E}}\cdot\hat{\mathbf{x}}$ coupling, and included, as an approximation necessary to causality, the extension of the range of integration to the whole real axis. This is an illustration of the Hegerfeldt theorem \cite{HegerfeldtFermi}, which states that non-causalities will always arise for Hamiltonians bounded by below. The relevant Hamiltonian here for the Hegerfeldt theorem is the Hamiltonian $\hat{H}_R$ (\ref{eq:Hamilton}) of the free field, which has $\mathbb{R}_+$ as its spectrum and is hence bounded. The non-causality seen in (\ref{eq:ThreePlusMinusMinus}) is of a different kind\textemdash it is not a manifestation of the Hegerfeldt theorem\textemdash and is not known in the literature. See sect.~\ref{sec:Dsc} for further discussion. Notice that the $n=3$ case at hand corresponds, as can be seen from (\ref{eq:DipoleGreen}), to the near-field part of the emitted photon wave function. In the present manuscript we will rather focus on the mid- and far-field contributions to the electric field, and shall return to the delicate question of the near-field in an upcoming manuscript.\\

We now focus on the Hegerfeldt noncausality. To investigate this particular point, we compute the convolutions of (\ref{eq:PlusMinusMinusY}) and (\ref{eq:ThreePlusMinusMinus}) with the principal value as prescribed by (\ref{eq:ConvWithDeltavpFuture}). In the mid- and far-field we get ($n\in\left\{1,2\right\}$)
\begin{widetext}
\begin{equation} \label{eq:NonC}
\begin{aligned} [b]
C_{n\left(\mathrm{dip}\right)}\left(\left|\left|\mathbf{x}\right|\right|,t\right)&\equiv-\frac{\mathrm{i}}{2\pi}\left[\mathrm{vp}\,\frac{1}{\cdot}*\bar{f}_{n\left(\mathrm{dip}\right)}\left(\cdot,t\right)\right]\left(\left|\left|\mathbf{x}\right|\right|\right)\\
&=\mathrm{e}^{-\mathrm{i}\frac{\Omega_0}{c}\left|\left|\mathbf{x}\right|\right|}\left(\frac{\Omega_0}{c}\right)^{2-n}\left[-\mathrm{Ei}\left(\mathrm{i}\frac{\Omega_0}{c}\left|\left|\mathbf{x}\right|\right|\right)+\mathrm{Ei}\left(\mathrm{i}\frac{\Omega_0}{c}\left(\left|\left|\mathbf{x}\right|\right|+ct\right)\right)\right]
\end{aligned}
\end{equation}
and in the near-field
\begin{equation} \label{eq:NonCZero}
\begin{aligned} [b]
C_{3\left(\mathrm{dip}\right)}\left(\left|\left|\mathbf{x}\right|\right|,t\right)=&\equiv-\frac{\mathrm{i}}{2\pi}\left[\mathrm{vp}\,\frac{1}{\cdot}*\bar{f}_{3\left(\mathrm{dip}\right)}\left(\cdot,t\right)\right]\left(\left|\left|\mathbf{x}\right|\right|\right)\\
&=\mathrm{e}^{-\mathrm{i}\frac{\Omega_0}{c}\left|\left|\mathbf{x}\right|\right|}\frac{c}{\Omega_0}\left[-\mathrm{Ei}\left(\mathrm{i}\frac{\Omega_0}{c}\left|\left|\mathbf{x}\right|\right|\right)+\mathrm{Ei}\left(\mathrm{i}\frac{\Omega_0}{c}\left(\left|\left|\mathbf{x}\right|\right|+ct\right)\right)\right]\\
&-\frac{1}{\Omega_0}\mathrm{vp}\left[\int_{-\left|\left|\mathbf{x}\right|\right|}^{+\infty}\frac{\mathrm{d}v}{v}-\mathrm{e}^{\mathrm{i}\Omega_0t}\int_{-\left|\left|\mathbf{x}\right|\right|-ct}^{+\infty}\frac{\mathrm{d}u}{u}\right]
\end{aligned}
\end{equation}
so that
\begin{equation} \label{eq:NonCZero}
\begin{aligned} [b]
C_{3\left(\mathrm{dip}\right)}\left(-\left|\left|\mathbf{x}\right|\right|,t\right)-C_{3\left(\mathrm{dip}\right)}\left(\left|\left|\mathbf{x}\right|\right|,t\right)&=\frac{c}{\Omega_0}\left[\mathrm{e}^{\mathrm{i}\frac{\Omega_0}{c}\left|\left|\mathbf{x}\right|\right|}\left[-\mathrm{Ei}\left(-\mathrm{i}\frac{\Omega_0}{c}\left|\left|\mathbf{x}\right|\right|\right)+\mathrm{Ei}\left(\mathrm{i}\frac{\Omega_0}{c}\left(-\left|\left|\mathbf{x}\right|\right|+ct\right)\right)\right]\right.\\
&\hspace{25pt}\left.-\mathrm{e}^{-\mathrm{i}\frac{\Omega_0}{c}\left|\left|\mathbf{x}\right|\right|}\left[-\mathrm{Ei}\left(\mathrm{i}\frac{\Omega_0}{c}\left|\left|\mathbf{x}\right|\right|\right)+\mathrm{Ei}\left(\mathrm{i}\frac{\Omega_0}{c}\left(\left|\left|\mathbf{x}\right|\right|+ct\right)\right)\right]\right]\\
&-\frac{1}{\Omega_0}\mathrm{e}^{\mathrm{i}\Omega_0t}\log\left(\frac{\left|\left|\mathbf{x}\right|\right|+ct}{\left|-\left|\left|\mathbf{x}\right|\right|+ct\right|}\right).
\end{aligned}
\end{equation}
\end{widetext}
Here $\mathrm{Ei}$ stands for the exponential integral \cite{AbramowitzStegun}
\begin{equation} \label{eq:ExpInt}
\mathrm{Ei}\left(x\right)\equiv-\int_{-x}^{+\infty}\mathrm{d}u\frac{\mathrm{e}^{-u}}{u}.
\end{equation}
Note that to deduce $\bar{f}_{n\left(\mathrm{dip}\right)}$, as given by (\ref{eq:FermiIntegral}), from (\ref{eq:ConvWithDeltavpFuture}) and (\ref{eq:Bob}), we made unwarranted use of Jordan's lemma: the integral of $f_{n\left(\mathrm{dip}\right)}$ over the semicircle from $\gamma_{\mathrm{top}}$ (see Fig.~\ref{fig:Camille}) is not zero for $n=1$ in the limit of large semicircle radius. On the contrary, this integral diverges in that limit \footnote{This was also true of the integrals $H_{n\left(\mathrm{std}\right)}^{\left(\pm\right)}$ for $n=1,2$ in the $\hat{\mathbf{E}}\cdot\hat{\mathbf{x}}$ case of the previous sect.~\ref{sec:CausalWW}.}. It is thus clear that the dipole approximation forbids a clear, consistent treatment of the problem at hand (unless a cutoff is introduced around the frequency $\omega_X=\left(3c\right)/\left(2a_0\right)$ as explained in \cite{EdouardIsa,FredZ}. However if a cutoff is introduced we cannot avoid Hegerfeldt-type non-causalities). In the next sect.~\ref{sec:MyHeart} we do away with this approximation. Nevertheless, we will see that very similar terms to those obtained here in the framework of the dipole approximation arise. Since the expressions obtained are much more involved in the next sect.~\ref{sec:MyHeart}, we study the less complicated results of the present section in some detail, and this knowledge will come in handy for later. Namely, taking the $k_{\mathrm{X}}\rightarrow+\infty$ limit of the results of sect.~\ref{sec:MyHeart} yields the results in the dipole approximation. This confirms that the exact coupling provides the correct regularisation for the dipole approximation (see \cite{EdouardIsa,FredZ}).\\

We plot in Fig.~\ref{fig:ApDip} the square moduli of the causal (or pole) (\ref{eq:PlusMinusMinusY}) and noncausal (or principal value) (\ref{eq:NonC}) contributions to the far- and mid-field parts of the emitted photon wave function.\\ 
\begin{figure*}[t,h]
\begin{center}
\subfloat{\includegraphics[width=1.02875\columnwidth]{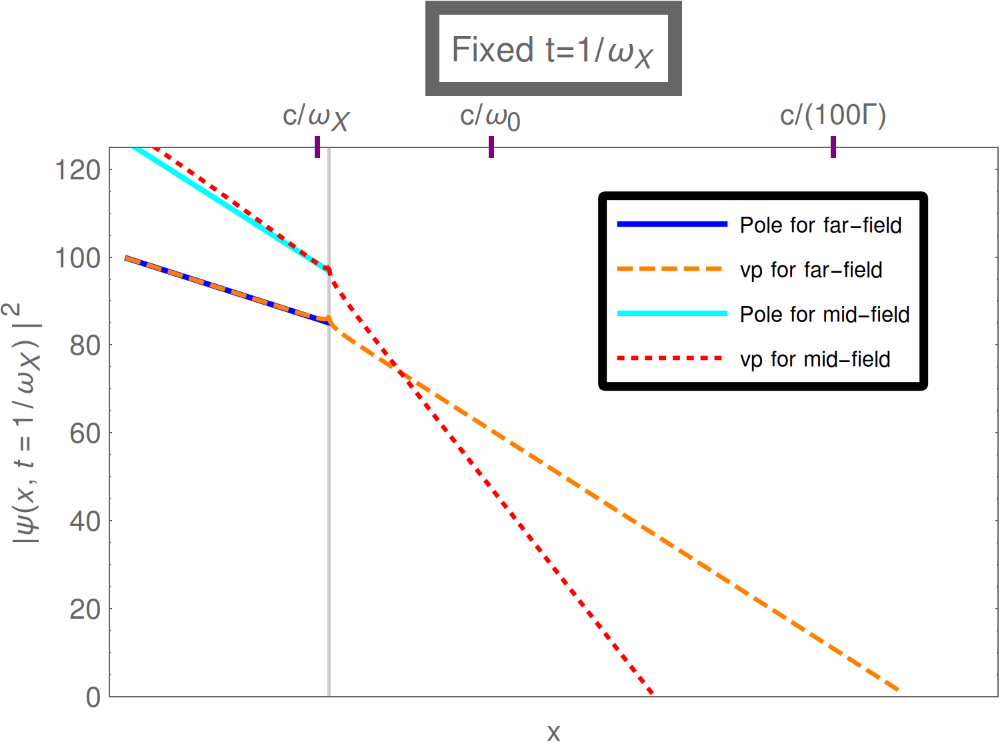}
\includegraphics[width=1.02875\columnwidth]{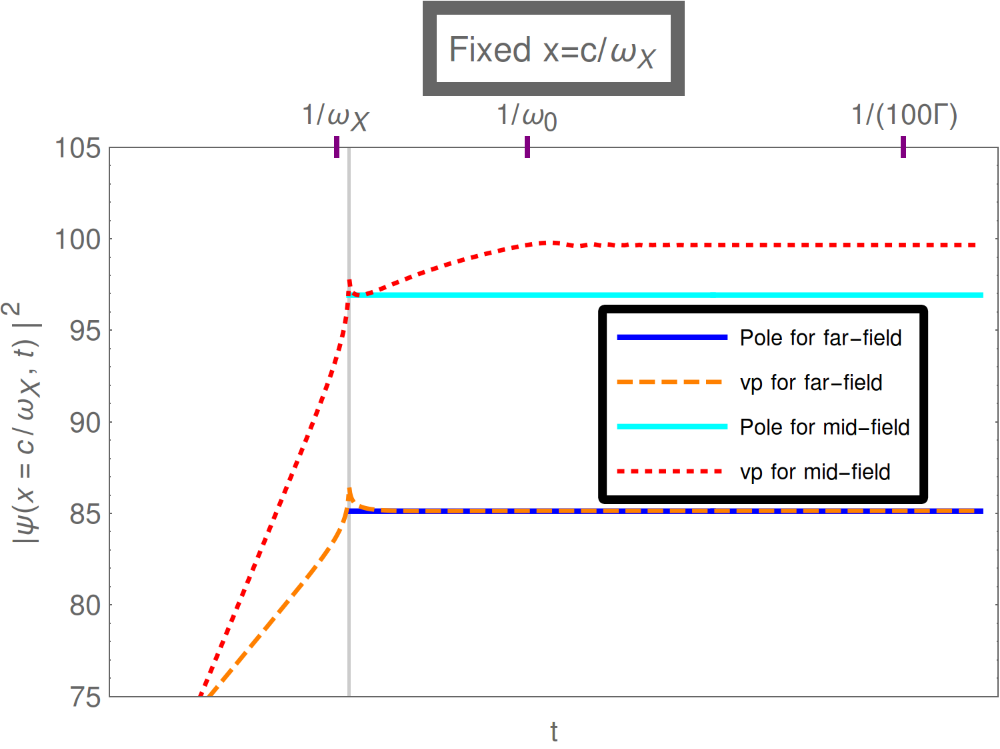}
}
\end{center}
%\caption{See caption below. \label{fig:ApDip}}
%\end{figure*}
%\begin{figure*}[t,h]
%\ContinuedFloat
\begin{center}
\subfloat{
\includegraphics[width=1.02875\columnwidth]{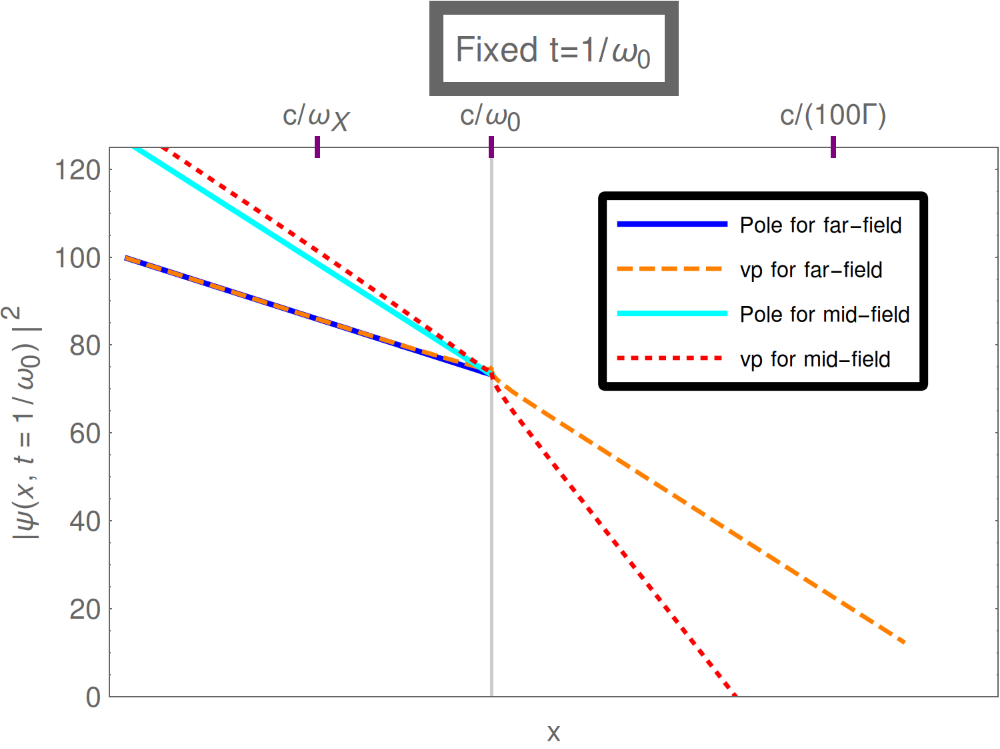}
\includegraphics[width=1.02875\columnwidth]{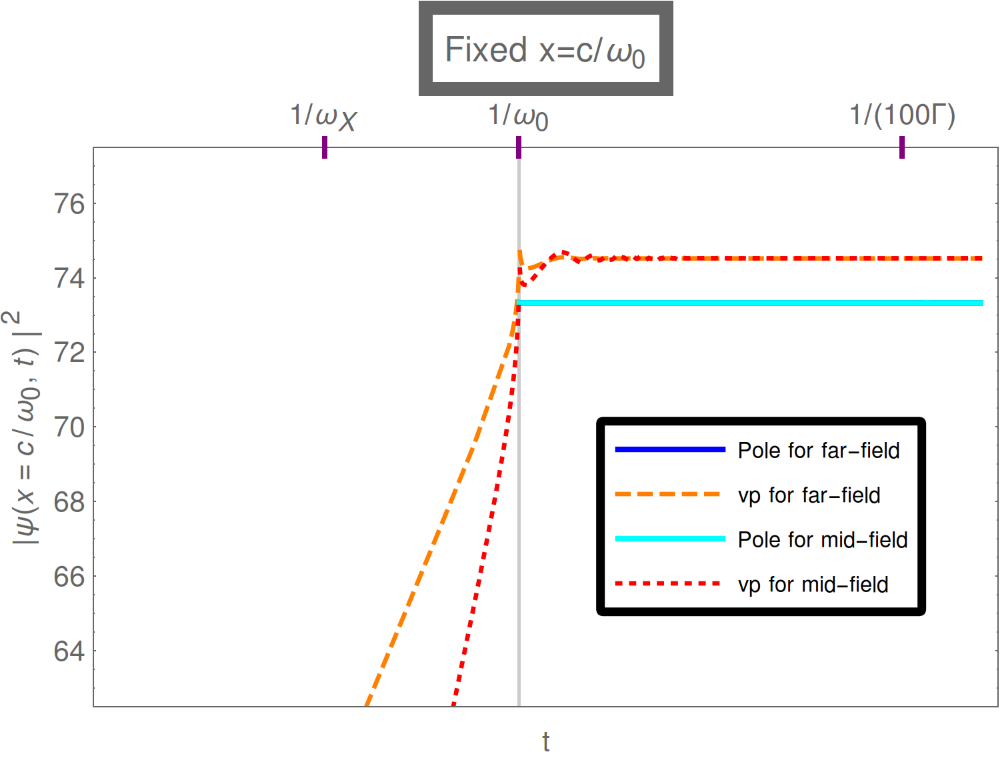}
}
\end{center}
\caption{See rest of figure and caption below. \label{fig:ApDip}}
\end{figure*}
\begin{figure*}[t,h]
\ContinuedFloat
\begin{center}
\subfloat{\includegraphics[width=1.02875\columnwidth]{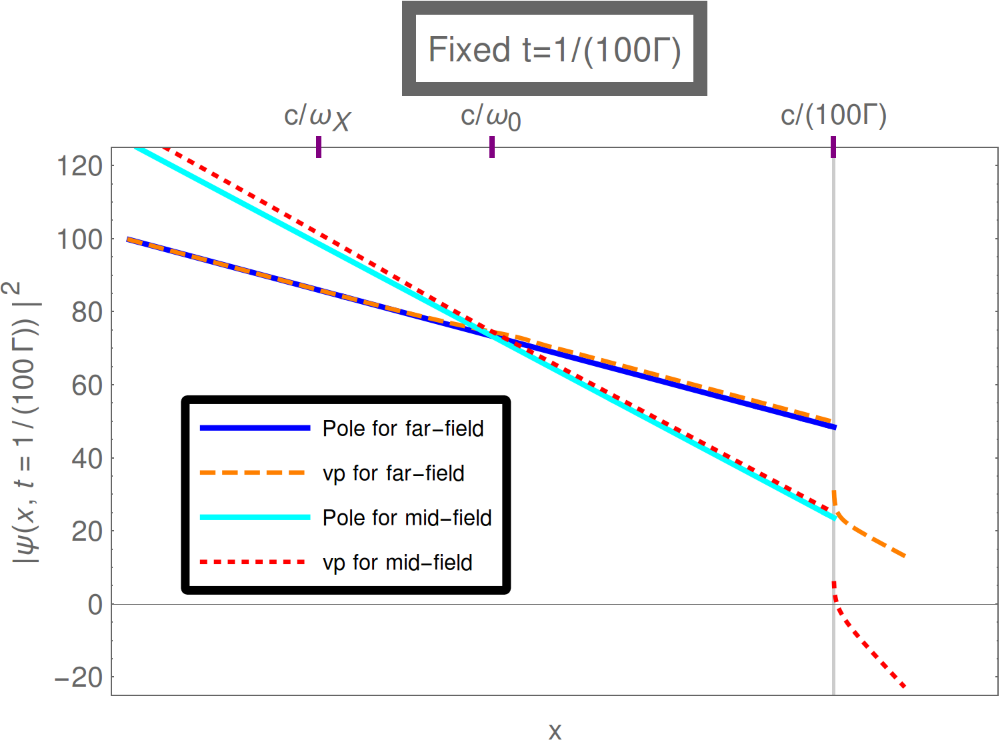}
\includegraphics[width=1.02875\columnwidth]{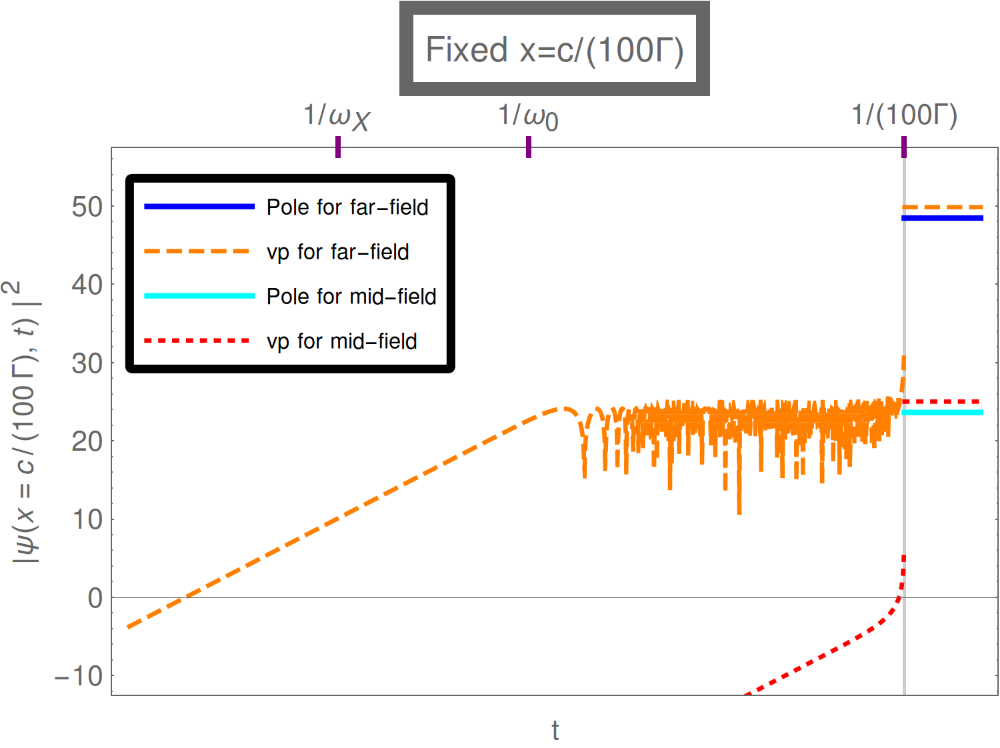}}
\end{center}
\caption{Square moduli of the contributions to the far- ($n=1$) and mid- ($n=2$) fields from the poles as given by (\ref{eq:PlusMinusMinusY}) and from the principal value integrals as given by (\ref{eq:NonC}) as a function of distance from the nucleus, for fixed time in the left column, and as a function of time, for fixed distance in the right column. This is for the minimal $\hat{\mathbf{A}}\cdot\hat{\mathbf{p}}$ coupling in the dipole approximation (sect.~\ref{subsec:AdotPDipHeg}). Both axes are logarithmic on all figures. The boundary of the lightcone is signalled by a vertical line. The reader will notice the noteworthy fact that for fixed $x=c/\omega_0$, the contributions from the mid-field and far-field are indistinguishable inside the lightcone. \label{fig:ApDip}}
\end{figure*}

Two noticeable patterns emerge: 
\begin{itemize}
\item{Inside the lightcone, we notice that the contributions to the far-field ($n=1$) from the pole term (\ref{eq:PlusMinusMinusY}) and from the principal value term (\ref{eq:NonC}) are almost indistinguishable, at least when $\left|\left|\mathbf{x}\right|\right|<c/\Omega_0$. This feature can be explained by simple asymptotic \cite{AbramowitzStegun} arguments: under the condition that we pick a spacetime point reasonably ``deep'' within the lightcone, we may make the approximation that $\left|\left|\mathbf{x}\right|\right|/\left(ct\right)\rightarrow0$ so that from (\ref{eq:NonC}) and the asymptotic series
\begin{equation} \label{eq:EiAsy}
\mathrm{Ei}\left(u\right)\underset{u\rightarrow+\infty}{\sim}\frac{\mathrm{e}^u}{u}\sum_{k=0}^{+\infty}\frac{k!}{u^k}
\end{equation}
we may write, with the extra help of the Taylor series for $\mathrm{Ei}\left(u\right)$ around $u=0$,
\begin{equation} \label{eq:LargeTAsy}
C_{1\left(\mathrm{dip}\right)}\left(-\left|\left|\mathbf{x}\right|\right|,t\right)-C_{1\left(\mathrm{dip}\right)}\left(\left|\left|\mathbf{x}\right|\right|,t\right)\underset{\frac{\Omega_0}{c}\left|\left|\mathbf{x}\right|\right|\rightarrow0}{\sim}\mathrm{i}\pi\frac{\Omega_0}{c},
\end{equation}
in agreement with the asymptotic behaviour of the causal part (\ref{eq:PlusMinusMinusY}) in the same $\left(\Omega_0/c\right)\left|\left|\mathbf{x}\right|\right|\ll1$ limit. This good agreement is not reached for the mid-field because the contributions from ingoing and outgoing waves are added instead of substracted, as seen in (\ref{eq:DipoleGreen}).
}
\item{Outside the lightcone (where the only contributions to the far-and mid-field, obviously, come from the principal value integrals), we notice on the graphs that the ($n=1$) far-field decays not as $1/\left|\left|\mathbf{x}\right|\right|$ as expected, but as $1/\left|\left|\mathbf{x}\right|\right|^2$. As for the ($n=2$) mid-field, it decays not as $1/\left|\left|\mathbf{x}\right|\right|^2$ as expected, but as $1/\left|\left|\mathbf{x}\right|\right|^4$. Again, this can be explained by the asymptotic behaviour \cite{AbramowitzStegun} of the exponential integral function: under the condition that we pick a spacetime point reasonably far away from the lightcone, we may make the approximation that $\left|\left|\mathbf{x}\right|\right|/\left(ct\right)\rightarrow+\infty$ so that from (\ref{eq:NonC}) and (\ref{eq:EiAsy}) we may write the asymptotic series
\begin{widetext}
\begin{subequations} \label{eq:LargeXAsy}
\begin{align}
C_{1\left(\mathrm{dip}\right)}\left(-\left|\left|\mathbf{x}\right|\right|,t\right)-C_{1\left(\mathrm{dip}\right)}\left(\left|\left|\mathbf{x}\right|\right|,t\right)&\simeq-\frac{2 \mathrm{i}\left(-1+\mathrm{e}^{\mathrm{i}\Omega_0t}\right)}{\left|\left|\mathbf{x}\right|\right|}\mbox{ (for large $\frac{\Omega_0}{c}\left|\left|\mathbf{x}\right|\right|$)},\\
C_{2\left(\mathrm{dip}\right)}\left(-\left|\left|\mathbf{x}\right|\right|,t\right)+C_{2\left(\mathrm{dip}\right)}\left(\left|\left|\mathbf{x}\right|\right|,t\right)&\simeq\frac{2 +\mathrm{e}^{\mathrm{i}\Omega_0t}\left(-2+2\mathrm{i}\Omega_0t\right)}{\left(\frac{\Omega_0}{c}\right)^2\left|\left|\mathbf{x}\right|\right|^2}\mbox{ (id.)}.
\end{align}
\end{subequations}
\end{widetext}
This accounts for the spacewise-decay behaviour described just above. The conclusion reached is interesting: outside the lightcone, the far-field decays more strongly than usual with increasing distance and mimics the usual behaviour of the mid-field ($1/\left|\left|\mathbf{x}\right|\right|^2$), while the mid-field decays much more strongly than usual with increasing distance, so that its $1/\left|\left|\mathbf{x}\right|\right|^4$ decay is stronger than the usual $1/\left|\left|\mathbf{x}\right|\right|^3$ decay of the near-field.}
\end{itemize}
Both these points are noteworthy features of the Hegerfeldt-noncausal terms, which come from the fact that the integration is restricted to positive electromagnetic frequencies, as it should. The first point confirms that the usual approximation consisting in extending the integration to the negative real semi-axis is fairly solid: inside the lightcone, we see (Fig.~\ref{fig:ApDip}) that the contribution from the (principal value) noncausal terms (\ref{eq:NonC}) just about equals that of the (pole) causal terms (\ref{eq:PlusMinusMinusY}), which means that the result yielded by extending the range of integration would be sensible. As for what happens outside the lightcone, we not only pointed the well-known fact that taking the absence of negative frequencies into account yields a nonzero result, but we noticed the interesting fact that the field decays more strongly with increasing distance than would be naively inferred from the usual behaviour of the mid- and far-field contributions to the emitted electric field. Keep in mind that the noncausal contributions to the emitted field are small, as seen on Fig.~\ref{fig:ApDip} (remember that the axes are logarithmic).

\section{Exact treatment for the minimal coupling} \label{sec:MyHeart}

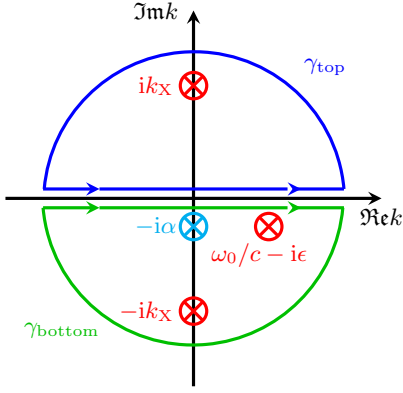
\begin{figure} [t]
\begin{center}
\begin{tikzpicture}[very thick, >=stealth]
\draw[->] (-2.5,0) -- (2.5,0);
\draw[->] (0,-2.5) -- (0,2.5);
\draw (2.5,-.25) node {\small $\mathfrak{Re}k$};
\draw (-.5,2.5) node {\small $\mathfrak{Im}k$};
\draw[red] (.875,-.5) -- (1.125,-.25);
\draw[red] (.875,-.25) -- (1.125,-.5);
\draw[red] (1,-.375) circle (5pt);
\draw[red] (.875,-.75) node {\small $\omega_0/c-\mathrm{i}\epsilon$};
\draw[cyan] (-.125,-.5) -- (.125,-.25);
\draw[cyan] (-.125,-.25) -- (.125,-.5);
\draw[cyan] (0,-.375) circle (5pt);
\draw[cyan] (-.5,-.375) node {\small $-\mathrm{i}\alpha$};
\draw[red] (-.125,1.625) -- (.125,1.375);
\draw[red] (-.125,1.375) -- (.125,1.625);
\draw[red] (0,1.5) circle (5pt);
\draw[red] (-.5,1.5) node {\small $\mathrm{i}k_{\mathrm{X}}$};
\draw[red] (-.125,-1.625) -- (.125,-1.375);
\draw[red] (-.125,-1.375) -- (.125,-1.625);
\draw[red] (0,-1.5) circle (5pt);
\draw[red] (-.625,-1.5) node {\small $-\mathrm{i}k_{\mathrm{X}}$};
\draw[blue, ->] (-2,.125) -- (-1.25,.125);
\draw[blue] (-1.25,.125) -- (1.25,.125);
\draw[blue, >-] (1.25,.125) -- (2,.125);
\draw[blue] (2,.125) arc (5:175:2);
\draw[blue] (1.75,1.75) node {\small $\gamma_{\mathrm{top}}$};
\draw[green!75!black, -<] (2,-.125) -- (1.25,-.125);
\draw[green!75!black] (1.25,-.125) -- (-1.25,-.125);
\draw[green!75!black, <-] (-1.25,-.125) -- (-2,-.125);
\draw[green!75!black] (-2,-.125) arc (185:355:2);
\draw[green!75!black] (-1.75,-1.75) node {\small $\gamma_{\mathrm{bottom}}$};
\end{tikzpicture}
\end{center}
\vspace{-10pt}
\caption{Jordan loops in the complex $k$-plane used to compute the Fourier transform (\ref{eq:TrueStart}). The (isolated) poles $\omega_0/c-\mathrm{i}\epsilon$ and $\pm\mathrm{i}k_{\mathrm{X}}$ are represented by red circled crosses while the (isolated) simple pole $-\mathrm{i}\alpha$ is represented by a cyan circled cross. \label{fig:CauchyPaths}}
\end{figure}

We now switch to a fully rigorous treatment of the problem. We use the minimal $\hat{\mathbf{A}}\cdot\hat{\mathbf{p}}$ form of the atom-field coupling. We no longer work in the dipole approximation, but shall instead use the exact interaction matrix element (\ref{eq:MatrixMultipole}). We also no longer perform the Wigner-Weisskopf approximation but use perturbation theory at short times \footnote{Note that it is very easy to show that if, instead of this time-dependent perturbation theory, we plug in the expression from the usual Wigner-Weisskopf approximation (which has been shown \cite{FacchiPascazio} to provide a very good approximation to the dynamics of the system (except at very short times), up to very small corrections), our results for the matrix elements of the electric field hold, with a grain of salt: $\omega_0$ should be replaced by $\omega_0+\omega_{\mathrm{LS}}-\left(\mathrm{i}/2\right)\Gamma$. Therefore, our perturbative treatment works directly at short times and indirectly (through the substitution which we just specified) at intermediate times. It only fails at very long times, where the decay becomes nonexponential \cite{FacchiPascazio}.}. Finally, we do not extend the range of integration to the negative real semi-axis as was done in sects.~\ref{sec:CausalWW} and \ref{subsec:LongPole}. In a time-dependent perturbative treatment of the present problem, we approximate (\ref{eq:PsiBeta}) to first order in time, which consists \cite{FredZ} in considering that $c_{\mathrm{e}}\left(t\right)=1$, so that the photon wave function is approximated by
\begin{widetext}
\begin{equation} \label{eq:MyStart}
\bm{\psi}_\perp\left(\mathbf{x},t\right)=-\frac{2^{\frac{7}{2}}}{3^4}\frac{\hbar e}{\epsilon_0m_ea_0}\int_0^{+\infty}\frac{\mathrm{d}k}{\left(2\pi\right)^2}\frac{k^2}{\left[1+\left(\frac{k}{k_{\mathrm{X}}}\right)^2\right]^2}\mathrm{e}^{-\mathrm{i}\omega_0t}
\mathbf{I}\left(k,\left|\left|\mathbf{x}\right|\right|\right)\frac{1-\mathrm{e}^{-\mathrm{i}\left(ck-\omega_0\right)t}}{ck-\omega_0}.
\end{equation}
We find ourselves computing integrals of the type
\begin{equation} \label{eq:ThisIsK}
H_n^{\left(\pm\right)}\left(\left|\left|\mathbf{x}\right|\right|,t\right)\equiv\int_0^{+\infty}\mathrm{d}k\,\frac{k^{2-n}}{\left[1+\left(\frac{k}{k_{\mathrm{X}}}\right)^2\right]^2}\mathrm{e}^{\pm\mathrm{i}k\left|\left|\mathbf{x}\right|\right|}\,\frac{1-\mathrm{e}^{-\mathrm{i}\left(ck-\omega_0\right)t}}{ck-\omega_0}
\end{equation}
\end{widetext}
with $n\in\left\{1,2,3\right\}$. The integrand is similar to that in (\ref{eq:Bob}), but features two extra poles at $k=\pm\mathrm{i}k_{\mathrm{X}}$, and its ``Wigner-Weisskopf'' pole at $k=\omega_0/c$ sits on the real axis. Since the latter is only an artificial singularity, we can shift it to the lower half plane as seen on Fig.~\ref{fig:CauchyPaths}. As was the case in the previous sect.~\ref{sec:PreStorm}, for $n=3$ the integrand has a singularity at $k=0$, which we are allowed to shift to $k=-\mathrm{i}\alpha$ since we will be interested in computing the difference $\left(H_3^{\left(+\right)}-H_3^{\left(-\right)}\right)\left(\left|\left|\mathbf{x}\right|\right|,t\right)$, which is an integral over a function which, as seen from (\ref{eq:ThisIsK}), has only an artificial singularity at $k=0$. We can rewrite (\ref{eq:ThisIsK}) as
\begin{widetext}
\begin{equation} \label{eq:ThisIsKRewritten}
\begin{aligned} [b]
H_n^{\left(\pm\right)}\left(\left|\left|\mathbf{x}\right|\right|,t\right)&\equiv\int_{-\infty}^{+\infty}\hspace{-7.5pt}\mathrm{d}k\,\frac{\theta\left(k\right)\left(k+\mathrm{i}\alpha\right)^{2-n}}{\left[1+\left(\frac{k}{k_{\mathrm{X}}}\right)^2\right]^2}\mathrm{e}^{\pm\mathrm{i}k\left|\left|\mathbf{x}\right|\right|}\,\frac{1-\mathrm{e}^{-\mathrm{i}\left(ck-\omega_0\right)t}}{ck-\left(\omega_0-\mathrm{i}c\,\epsilon\right)}\\
&\equiv\int_{-\infty}^{+\infty}\mathrm{d}k\,\theta\left(k\right)\,f_n\left(k,t\right)\mathrm{e}^{\pm\mathrm{i}k\left|\left|\mathbf{x}\right|\right|}\\
&\equiv\int_{-\infty}^{+\infty}\mathrm{d}k\,\theta\left(k\right)\left(g_n\left(k\right)-h_n\left(k,t\right)\right)\mathrm{e}^{\pm\mathrm{i}k\left|\left|\mathbf{x}\right|\right|}
\end{aligned}
\end{equation}
\end{widetext}
where it is implied that the limit $\alpha\rightarrow0^+$, $\epsilon\rightarrow0^+$ should be taken outside the integral. Here the functions $g_n$ and $h_n$ read
{\allowdisplaybreaks
\begin{subequations} \label{eq:GandH}
\begin{align}
g_n\left(k\right)&\equiv\frac{\left(k+\mathrm{i}\alpha\right)^{2-n}}{\left[1+\left(\frac{k}{k_{\mathrm{X}}}\right)^2\right]^2}\,\frac{1}{ck-\left(\omega_0-\mathrm{i}c\,\epsilon\right)},\\
h_n\left(k,t\right)&\equiv\frac{\left(k+\mathrm{i}\alpha\right)^{2-n}}{\left[1+\left(\frac{k}{k_{\mathrm{X}}}\right)^2\right]^2}\,\frac{\mathrm{e}^{-\mathrm{i}\left(ck-\omega_0\right)t}}{ck-\left(\omega_0-\mathrm{i}c\,\epsilon\right)}.
\end{align}
\end{subequations}
}
Once again we use (\ref{eq:ConvWithDeltavpFuture}) and hence need to compute the Fourier transform
\begin{equation} \label{eq:TrueStart}
\bar{f}_n\left(\left|\left|\mathbf{x}\right|\right|,t\right)=\int_{-\infty}^{+\infty}\mathrm{d}k\,f_n\left(k,t\right)\mathrm{e}^{-\mathrm{i}kx}
\end{equation}
of $f_n$. We use Cauchy's residue theorem. We know from (\ref{eq:ThisIsK}) that $f_n$ has a first order pole at $\omega_0/c-\mathrm{i}\epsilon$ and two second order poles at $\pm\mathrm{i}k_{\mathrm{X}}$, pictured on Fig.~\ref{fig:CauchyPaths}. From (\ref{eq:ThisIsKRewritten}) and (\ref{eq:TrueStart}) we see that we have to close the integration path (Jordan loop) in the lower half of the complex plane for $\left|\left|\mathbf{x}\right|\right|>0$ and $\left|\left|\mathbf{x}\right|\right|+ct>0$ for $g_n$ and $h_n$ respectively, and in the upper half of the plane for $\left|\left|\mathbf{x}\right|\right|<0$ or $\left|\left|\mathbf{x}\right|\right|+ct<0$ for $g_n$ and $h_n$ respectively.\\

\begin{widetext}
It can be checked that the residues of $g_n\left(\omega\right)\mathrm{e}^{-\mathrm{i}kx}$ and $h_n\left(\omega\right)\mathrm{e}^{-\mathrm{i}kx}$ read
\begin{subequations} \label{eq:GResidues}
\begin{align}
\mathrm{Res}\left(g_n\left(\cdot\right)\mathrm{e}^{-\mathrm{i}\cdot\left|\left|\mathbf{x}\right|\right|},\frac{\omega_0}{c}-\mathrm{i}\epsilon\right)&\underset{\epsilon\rightarrow0^+}{\longrightarrow}\mathrm{e}^{-\mathrm{i}\frac{\omega_0}{c}\left|\left|\mathbf{x}\right|\right|}\,G_0^{\left(n\right)},\\
\mathrm{Res}\left(g_n\left(\cdot\right)\mathrm{e}^{-\mathrm{i}\cdot\left|\left|\mathbf{x}\right|\right|},\mathrm{i}k_{\mathrm{X}}\right)&\underset{\epsilon\rightarrow0^+}{\longrightarrow}\mathrm{e}^{k_{\mathrm{X}}\left|\left|\mathbf{x}\right|\right|}\left(\gamma_0^{+\left(n\right)}+\gamma_1^{+\left(n\right)}\,\left|\left|\mathbf{x}\right|\right|\right),\\
\mathrm{Res}\left(g_n\left(\cdot\right)\mathrm{e}^{-\mathrm{i}\cdot\left|\left|\mathbf{x}\right|\right|},-\mathrm{i}k_{\mathrm{X}}\right)&\underset{\epsilon\rightarrow0^+}{\longrightarrow}\mathrm{e}^{-k_{\mathrm{X}}\left|\left|\mathbf{x}\right|\right|}\left(\gamma_0^{-\left(n\right)}+\gamma_1^{-\left(n\right)}\,\left|\left|\mathbf{x}\right|\right|\right)
\end{align}
\end{subequations}
and
\begin{subequations} \label{eq:HResidues}
\begin{align}
\mathrm{Res}\left(h_n\left(\cdot,t\right)\mathrm{e}^{-\mathrm{i}\cdot\left|\left|\mathbf{x}\right|\right|},\frac{\omega_0}{c}-\mathrm{i}\epsilon\right)&\underset{\epsilon\rightarrow0^+}{\longrightarrow}\mathrm{e}^{-\mathrm{i}\frac{\omega_0}{c}\left|\left|\mathbf{x}\right|\right|}\,G_0^{\left(n\right)},\\
\mathrm{Res}\left(h_n\left(\cdot,t\right)\mathrm{e}^{-\mathrm{i}\cdot\left|\left|\mathbf{x}\right|\right|},\mathrm{i}k_{\mathrm{X}}\right)&\underset{\epsilon\rightarrow0^+}{\longrightarrow}\mathrm{e}^{k_{\mathrm{X}}\left|\left|\mathbf{x}\right|\right|}\left(\gamma_0^{+\left(n\right)}+\gamma_1^{+\left(n\right)}\left(\left|\left|\mathbf{x}\right|\right|+ct\right)\right),\\
\mathrm{Res}\left(h_n\left(\cdot,t\right)\mathrm{e}^{-\mathrm{i}\cdot\left|\left|\mathbf{x}\right|\right|},-\mathrm{i}k_{\mathrm{X}}\right)&\underset{\epsilon\rightarrow0^+}{\longrightarrow}\mathrm{e}^{-k_{\mathrm{X}}\left|\left|\mathbf{x}\right|\right|}\left(\gamma_0^{-\left(n\right)}+\gamma_1^{-\left(n\right)}\left(\left|\left|\mathbf{x}\right|\right|+ct\right)\right)
\end{align}
\end{subequations}
\end{widetext}
where the $G_0^{\left(n\right)}$ and $\gamma_i^{\pm\left(n\right)}$ depend on $n$, as suggested by the notation. One can see that
\begin{align*}
\gamma_0^{+\left(n\right)}=\gamma_0^{-*\left(n\right)}&\equiv\gamma_0^{\left(n\right)},\\
\gamma_1^{+\left(n\right)}=-\gamma_1^{-*\left(n\right)}&\equiv\gamma_1^{\left(n\right)}.
\end{align*}
We give
\begin{subequations} \label{eq:ResidueGalore}
\begin{align}
G_0^{\left(n\right)}&=\frac{\left(\frac{\omega_0}{c}\right)^{2-n}\left(ck_{\mathrm{X}}\right)^4}{\left(\omega_0^2+c^2k_{\mathrm{X}}^2\right)^2},\\
\gamma_0^{\left(n\right)}&=\frac{ck_{\mathrm{X}}\left(\mathrm{i}k_{\mathrm{X}}\right)^{2-n}\left[-\mathrm{i}\left(-3+n\right)\omega_0-\left(n-2\right)ck_{\mathrm{X}}\right]}{4\left(\mathrm{i}\omega_0+ck_{\mathrm{X}}\right)^2},\\
\gamma_1^{\left(n\right)}&=\frac{ck_{\mathrm{X}}^2\left(\mathrm{i}k_{\mathrm{X}}\right)^{2-n}}{4\left(\mathrm{i}\omega_0+ck_{\mathrm{X}}\right)}.
\end{align}
\end{subequations}
The Fourier transform (\ref{eq:TrueStart}) is thus given for $n\in\left\{1,2\right\}$ by
\begin{widetext}
\begin{equation} \label{eq:CplxResidue}
\begin{aligned} [b]
\frac{\bar{f}_n\left(\left|\left|\mathbf{x}\right|\right|,t\right)}{2\mathrm{i}\pi}&=-\theta\left(\left|\left|\mathbf{x}\right|\right|\right)\left[\mathrm{e}^{-k_{\mathrm{X}}\left|\left|\mathbf{x}\right|\right|}\left(\gamma_0^{*\left(n\right)}-\gamma_1^{*\left(n\right)}\left|\left|\mathbf{x}\right|\right|\right)+\mathrm{e}^{-\mathrm{i}\frac{\omega_0}{c}\left|\left|\mathbf{x}\right|\right|}G_0^{\left(n\right)}\right]+\theta\left(-\left|\left|\mathbf{x}\right|\right|\right)\left[\mathrm{e}^{-k_{\mathrm{X}}\left|\left|\mathbf{x}\right|\right|}\left(\gamma_0^{\left(n\right)}+\gamma_1^{\left(n\right)}\left|\left|\mathbf{x}\right|\right|\right)\right]\\
&+\theta\left(\left|\left|\mathbf{x}\right|\right|+ct\right)\left[\mathrm{e}^{\mathrm{i}\omega_0t}\mathrm{e}^{-k_{\mathrm{X}}\left(\left|\left|\mathbf{x}\right|\right|+ct\right)}\left(\gamma_0^{*\left(n\right)}-\gamma_1^{*\left(n\right)}\left(\left|\left|\mathbf{x}\right|\right|+ct\right)\right)+\mathrm{e}^{-\mathrm{i}\frac{\omega_0}{c}\left|\left|\mathbf{x}\right|\right|}G_0^{\left(n\right)}\right]\\
&-\theta\left(-\left(\left|\left|\mathbf{x}\right|\right|+ct\right)\right)\left[\mathrm{e}^{\mathrm{i}\omega_0t}\mathrm{e}^{k_{\mathrm{X}}\left(\left|\left|\mathbf{x}\right|\right|+ct\right)}\left(\gamma_0^{\left(n\right)}+\gamma_1^{\left(n\right)}\left(\left|\left|\mathbf{x}\right|\right|+ct\right)\right)\right].
\end{aligned}
\end{equation}
We can then compute the convolution\textemdash which we call $C_n\left(\left|\left|\mathbf{x}\right|\right|,t\right)$\textemdash with the principal value as prescribed by (\ref{eq:ConvWithDeltavpFuture}). It yields, still for $n\in\left\{1,2\right\}$,
\begin{equation} \label{eq:TrustMeMaybeX}
\begin{aligned} [b]
C_n\left(\left|\left|\mathbf{x}\right|\right|,t\right)&\equiv-\frac{\mathrm{i}}{2\pi}\left[\mathrm{vp}\,\frac{1}{\cdot}*\bar{f}_n\left(\cdot,t\right)\right]\left(\left|\left|\mathbf{x}\right|\right|\right)\\
&=\mathrm{e}^{-k_{\mathrm{X}}\left|\left|\mathbf{x}\right|\right|}\left[\left(\gamma_0^{*\left(n\right)}-\gamma_1^{*\left(n\right)}\left|\left|\mathbf{x}\right|\right|\right)\left[-\mathrm{Ei}\left(k_{\mathrm{X}}\left|\left|\mathbf{x}\right|\right|\right)+\mathrm{e}^{\left(\mathrm{i}\omega_0-ck_{\mathrm{X}}\right)t}\mathrm{Ei}\left(k_{\mathrm{X}}\left(\left|\left|\mathbf{x}\right|\right|+ct\right)\right)\right]\right.\\
&\hspace{32.5pt}\left.+\frac{\gamma_1^{*\left(n\right)}}{k_{\mathrm{X}}}\left(-\mathrm{e}^{k_{\mathrm{X}}\left|\left|\mathbf{x}\right|\right|}+\mathrm{e}^{\left(\mathrm{i}\omega_0-ck_{\mathrm{X}}\right)t}\mathrm{e}^{k_{\mathrm{X}}\left(\left|\left|\mathbf{x}\right|\right|+ct\right)}\right)-c\gamma_1^{*\left(n\right)}t\,\mathrm{e}^{\left(\mathrm{i}\omega_0-ck_{\mathrm{X}}\right)t}\mathrm{Ei}\left(k_{\mathrm{X}}\left(\left|\left|\mathbf{x}\right|\right|+ct\right)\right)\right]\\
&+\mathrm{e}^{k_{\mathrm{X}}\left|\left|\mathbf{x}\right|\right|}\left[\left(\gamma_0^{\left(n\right)}+\gamma_1^{\left(n\right)}\left|\left|\mathbf{x}\right|\right|\right)\left[-\mathrm{Ei}\left(-k_{\mathrm{X}}\left|\left|\mathbf{x}\right|\right|\right)+\mathrm{e}^{\left(\mathrm{i}\omega_0+ck_{\mathrm{X}}\right)t}\mathrm{Ei}\left(k_{\mathrm{X}}\left(-\left|\left|\mathbf{x}\right|\right|-ct\right)\right)\right]\right.\\
&\hspace{32.5pt}\left.+\frac{\gamma_1^{\left(n\right)}}{k_{\mathrm{X}}}\left(-\mathrm{e}^{-k_{\mathrm{X}}\left|\left|\mathbf{x}\right|\right|}+\mathrm{e}^{\left(\mathrm{i}\omega_0+ck_{\mathrm{X}}\right)t}\mathrm{e}^{k_{\mathrm{X}}\left(-\left|\left|\mathbf{x}\right|\right|-ct\right)}\right)+c\gamma_1^{\left(n\right)}t\,\mathrm{e}^{\left(\mathrm{i}\omega_0+ck_{\mathrm{X}}\right)t}\mathrm{Ei}\left(k_{\mathrm{X}}\left(-\left|\left|\mathbf{x}\right|\right|-ct\right)\right)\right]\\
&+\mathrm{e}^{-\mathrm{i}\frac{\omega_0}{c}\left|\left|\mathbf{x}\right|\right|}\,G_0^{\left(n\right)}\left[-\mathrm{Ei}\left(\mathrm{i}\frac{\omega_0}{c}\left|\left|\mathbf{x}\right|\right|\right)+\mathrm{Ei}\left(\mathrm{i}\frac{\omega_0}{c}\left(\left|\left|\mathbf{x}\right|\right|+ct\right)\right)\right].
\end{aligned}
\end{equation}
Now, remember that for $n=3$, the integrand in (\ref{eq:ThisIsK}) has a (simple) pole at $k=0$. As argued above (\ref{eq:ThisIsKRewritten}), we can shift this singularity away from the real axis $0\rightarrow-\mathrm{i}\alpha$ to compute the integral (see Fig.~\ref{fig:CauchyPaths}), before taking the limit $\alpha\rightarrow0$ at the end. This yields
\begin{equation} \label{eq:CplxResidueZeroX}
\begin{aligned} [b]
\frac{\bar{f}_3\left(\left|\left|\mathbf{x}\right|\right|,t\right)}{2\mathrm{i}\pi}=&-\theta\left(\left|\left|\mathbf{x}\right|\right|\right)\left[\mathrm{e}^{-k_{\mathrm{X}}\left|\left|\mathbf{x}\right|\right|}\left(\gamma_0^{*\left(3\right)}-\gamma_1^{*\left(3\right)}\left|\left|\mathbf{x}\right|\right|\right)+\mathrm{e}^{-\mathrm{i}\frac{\omega_0}{c}\left|\left|\mathbf{x}\right|\right|}G_0^{\left(3\right)}\right]+\theta\left(-\left|\left|\mathbf{x}\right|\right|\right)\left[\mathrm{e}^{-k_{\mathrm{X}}\left|\left|\mathbf{x}\right|\right|}\left(\gamma_0^{\left(3\right)}+\gamma_1^{\left(3\right)}\left|\left|\mathbf{x}\right|\right|\right)\right]\\
&+\theta\left(\left|\left|\mathbf{x}\right|\right|+ct\right)\left[\mathrm{e}^{\mathrm{i}\omega_0t}\mathrm{e}^{-k_{\mathrm{X}}\left(\left|\left|\mathbf{x}\right|\right|+ct\right)}\left(\gamma_0^{*\left(3\right)}-\gamma_1^{*\left(3\right)}\left(\left|\left|\mathbf{x}\right|\right|+ct\right)\right)+\mathrm{e}^{-\mathrm{i}\frac{\omega_0}{c}\left|\left|\mathbf{x}\right|\right|}G_0^{\left(3\right)}\right]\\
&-\theta\left(-\left(\left|\left|\mathbf{x}\right|\right|+ct\right)\right)\left[\mathrm{e}^{\mathrm{i}\omega_0t}\mathrm{e}^{k_{\mathrm{X}}\left(\left|\left|\mathbf{x}\right|\right|+ct\right)}\left(\gamma_0^{\left(3\right)}+\gamma_1^{\left(3\right)}\left(\left|\left|\mathbf{x}\right|\right|+ct\right)\right)\right]+\frac{1}{\omega_0}\left[\theta\left(\left|\left|\mathbf{x}\right|\right|\right)-\theta\left(\left|\left|\mathbf{x}\right|\right|+ct\right)\mathrm{e}^{\mathrm{i}\omega_0t}\right]
\end{aligned}
\end{equation}
and
\begin{equation} \label{eq:TrustMeMaybeZeroX}
\begin{aligned} [b]
C_3\left(\left|\left|\mathbf{x}\right|\right|,t\right)=&\mathrm{e}^{-k_{\mathrm{X}}\left|\left|\mathbf{x}\right|\right|}\left[\left(\gamma_0^{*\left(3\right)}-\gamma_1^{*\left(3\right)}\left|\left|\mathbf{x}\right|\right|\right)\left[-\mathrm{Ei}\left(k_{\mathrm{X}}\left|\left|\mathbf{x}\right|\right|\right)+\mathrm{e}^{\left(\mathrm{i}\omega_0-ck_{\mathrm{X}}\right)t}\mathrm{Ei}\left(k_{\mathrm{X}}\left(\left|\left|\mathbf{x}\right|\right|+ct\right)\right)\right]\right.\\
&\hspace{32.5pt}\left.+\frac{\gamma_1^{*\left(3\right)}}{k_{\mathrm{X}}}\left(-\mathrm{e}^{k_{\mathrm{X}}\left|\left|\mathbf{x}\right|\right|}+\mathrm{e}^{\left(\mathrm{i}\omega_0-ck_{\mathrm{X}}\right)t}\mathrm{e}^{k_{\mathrm{X}}\left(\left|\left|\mathbf{x}\right|\right|+ct\right)}\right)-c\gamma_1^{*\left(3\right)}t\,\mathrm{e}^{\left(\mathrm{i}\omega_0-ck_{\mathrm{X}}\right)t}\mathrm{Ei}\left(k_{\mathrm{X}}\left(\left|\left|\mathbf{x}\right|\right|+ct\right)\right)\right]\\
&+\mathrm{e}^{k_{\mathrm{X}}\left|\left|\mathbf{x}\right|\right|}\left[\left(\gamma_0^{\left(3\right)}+\gamma_1^{\left(3\right)}\left|\left|\mathbf{x}\right|\right|\right)\left[-\mathrm{Ei}\left(-k_{\mathrm{X}}\left|\left|\mathbf{x}\right|\right|\right)+\mathrm{e}^{\left(\mathrm{i}\omega_0+ck_{\mathrm{X}}\right)t}\mathrm{Ei}\left(k_{\mathrm{X}}\left(-\left|\left|\mathbf{x}\right|\right|-ct\right)\right)\right]\right.\\
&\hspace{32.5pt}\left.+\frac{\gamma_1^{\left(3\right)}}{k_{\mathrm{X}}}\left(-\mathrm{e}^{-k_{\mathrm{X}}\left|\left|\mathbf{x}\right|\right|}+\mathrm{e}^{\left(\mathrm{i}\omega_0+ck_{\mathrm{X}}\right)t}\mathrm{e}^{k_{\mathrm{X}}\left(-\left|\left|\mathbf{x}\right|\right|-ct\right)}\right)+c\gamma_1^{\left(3\right)}t\,\mathrm{e}^{\left(\mathrm{i}\omega_0+ck_{\mathrm{X}}\right)t}\mathrm{Ei}\left(k_{\mathrm{X}}\left(-\left|\left|\mathbf{x}\right|\right|-ct\right)\right)\right]\\
&+\mathrm{e}^{-\mathrm{i}\frac{\omega_0}{c}\left|\left|\mathbf{x}\right|\right|}\,G_0^{\left(3\right)}\left[-\mathrm{Ei}\left(\mathrm{i}\frac{\omega_0}{c}\left|\left|\mathbf{x}\right|\right|\right)+\mathrm{Ei}\left(\mathrm{i}\frac{\omega_0}{c}\left(\left|\left|\mathbf{x}\right|\right|+ct\right)\right)\right]-\frac{1}{\omega_0}\mathrm{vp}\left[\int_{-\left|\left|\mathbf{x}\right|\right|}^{+\infty}\frac{\mathrm{d}v}{v}-\mathrm{e}^{\mathrm{i}\omega_0t}\int_{-\left|\left|\mathbf{x}\right|\right|-ct}^{+\infty}\frac{\mathrm{d}u}{u}\right]
\end{aligned}
\end{equation}
which is an infinite quantity, but this is not a problem as we are interested in $C_3\left(-\left|\left|\mathbf{x}\right|\right|,t\right)-C_3\left(\left|\left|\mathbf{x}\right|\right|,t\right)$, which, as we shall see, is finite. Keeping in mind that $\left|\left|\mathbf{x}\right|\right|$ and $t$ are both positive we compute, for $n\in\left\{1,2\right\}$
\begin{equation} \label{eq:PlusMinusMinusX}
\begin{aligned} [b]
\bar{f}_n\left(-\left|\left|\mathbf{x}\right|\right|,t\right)\mp\bar{f}_n\left(\left|\left|\mathbf{x}\right|\right|,t\right)&=2\mathrm{i}\pi\left\{G_0^{\left(n\right)}\theta\left(-\left|\left|\mathbf{x}\right|\right|+ct\right)\mathrm{e}^{\mathrm{i}\frac{\omega_0}{c}\left|\left|\mathbf{x}\right|\right|}\right.\\
&\hspace{25pt}\left.+\theta\left(-\left|\left|\mathbf{x}\right|\right|+ct\right)\mathrm{e}^{\mathrm{i}\omega_0t}\left[\mathrm{e}^{-k_{\mathrm{X}}\left(ct-\left|\left|\mathbf{x}\right|\right|\right)}\left(\gamma_0^{*\left(n\right)}+\gamma_1^{*\left(n\right)}\left(\left|\left|\mathbf{x}\right|\right|-ct\right)\right)\right.\right.\\
&\hspace{112.5pt}\left.\left.+\mathrm{e}^{k_{\mathrm{X}}\left(ct-\left|\left|\mathbf{x}\right|\right|\right)}\left(\gamma_0^{\left(n\right)}-\gamma_1^{\left(n\right)}\left(\left|\left|\mathbf{x}\right|\right|-ct\right)\right)\right]\right.\\
&\hspace{25pt}\left.+\mathrm{e}^{-k_{\mathrm{X}}\left|\left|\mathbf{x}\right|\right|}\left[\left(\gamma_0^{\left(n\right)}\mp\gamma_0^{*\left(n\right)}\right)-\left(\gamma_1^{\left(n\right)}\mp\gamma_1^{*\left(n\right)}\right)\left|\left|\mathbf{x}\right|\right|\right]\right.\\
&\hspace{25pt}\left.-\mathrm{e}^{\mathrm{i}\omega_0t}\left[\mp\mathrm{e}^{-k_{\mathrm{X}}\left(ct+\left|\left|\mathbf{x}\right|\right|\right)}\left(\gamma_0^{*\left(n\right)}-\gamma_1^{*\left(n\right)}\left(\left|\left|\mathbf{x}\right|\right|+ct\right)\right)+\mathrm{e}^{k_{\mathrm{X}}\left(ct-\left|\left|\mathbf{x}\right|\right|\right)}\left(\gamma_0^{\left(n\right)}-\gamma_1^{\left(n\right)}\left(\left|\left|\mathbf{x}\right|\right|-ct\right)\right)\right]\right\}
\end{aligned}
\end{equation}
and, for $n=3$, the difference
\begin{equation} \label{eq:ThreePlusMinusMinusX}
\begin{aligned} [b]
\bar{f}_3\left(-\left|\left|\mathbf{x}\right|\right|,t\right)-\bar{f}_3\left(\left|\left|\mathbf{x}\right|\right|,t\right)&=2\mathrm{i}\pi\left\{G_0^{\left(3\right)}\theta\left(-\left|\left|\mathbf{x}\right|\right|+ct\right)\mathrm{e}^{\mathrm{i}\frac{\omega_0}{c}\left|\left|\mathbf{x}\right|\right|}\right.\\
&\hspace{25pt}\left.+\theta\left(-\left|\left|\mathbf{x}\right|\right|+ct\right)\mathrm{e}^{\mathrm{i}\omega_0t}\left[\mathrm{e}^{-k_{\mathrm{X}}\left(ct-\left|\left|\mathbf{x}\right|\right|\right)}\left(\gamma_0^{*\left(3\right)}+\gamma_1^{*\left(3\right)}\left(\left|\left|\mathbf{x}\right|\right|-ct\right)\right)\right.\right.\\
&\hspace{112.5pt}\left.\left.+\mathrm{e}^{k_{\mathrm{X}}\left(ct-\left|\left|\mathbf{x}\right|\right|\right)}\left(\gamma_0^{\left(3\right)}-\gamma_1^{\left(3\right)}\left(\left|\left|\mathbf{x}\right|\right|-ct\right)\right)\right]\right.\\
&\hspace{25pt}\left.+\mathrm{e}^{-k_{\mathrm{X}}\left|\left|\mathbf{x}\right|\right|}\left[\left(\gamma_0^{\left(3\right)}\mp\gamma_0^{*\left(3\right)}\right)-\left(\gamma_1^{\left(3\right)}\mp\gamma_1^{*\left(3\right)}\right)\left|\left|\mathbf{x}\right|\right|\right]\right.\\
&\hspace{25pt}\left.-\mathrm{e}^{\mathrm{i}\omega_0t}\left[\mp\mathrm{e}^{-k_{\mathrm{X}}\left(ct+\left|\left|\mathbf{x}\right|\right|\right)}\left(\gamma_0^{*\left(3\right)}-\gamma_1^{*\left(3\right)}\left(\left|\left|\mathbf{x}\right|\right|+ct\right)\right)+\mathrm{e}^{k_{\mathrm{X}}\left(ct-\left|\left|\mathbf{x}\right|\right|\right)}\left(\gamma_0^{\left(3\right)}-\gamma_1^{\left(3\right)}\left(\left|\left|\mathbf{x}\right|\right|-ct\right)\right)\right]\right.\\
&\hspace{25pt}\left.-\frac{1}{\omega_0}\left[1-\left(1-\theta\left(ct-\left|\left|\mathbf{x}\right|\right|\right)\right)\mathrm{e}^{\mathrm{i}\omega_0t}\right]\right\}.
\end{aligned}
\end{equation}
The quantities $C_n\left(-\left|\left|\mathbf{x}\right|\right|,t\right)\mp C_n\left(\left|\left|\mathbf{x}\right|\right|,t\right)$ are not illuminating enough to warrant their explicit writing out here, but their expression follows immediately from (\ref{eq:TrustMeMaybeX}) and (\ref{eq:TrustMeMaybeZeroX}). It is noteworthy, though, that the contribution to $C_3\left(-\left|\left|\mathbf{x}\right|\right|,t\right)\mp C_3\left(\left|\left|\mathbf{x}\right|\right|,t\right)$ (near-field) from the last summand on the right-hand side of (\ref{eq:TrustMeMaybeZeroX}) (the summand featuring the two principal value integrals) reads
\begin{equation} \label{eq:TrustMeMaybePlusMinusMinusMinusX}
-\frac{1}{\omega_0}\mathrm{e}^{\mathrm{i}\omega_0t}\log\left(\frac{\left|\left|\mathbf{x}\right|\right|+ct}{\left|-\left|\left|\mathbf{x}\right|\right|+ct\right|}\right).
\end{equation}
This particular result is reminiscent of the findings of Karpov \emph{et al.} in \cite{KarpovCausal} on the consequences that restricting the spectrum to positive frequencies has on localisation and causality. The cited work \cite{KarpovCausal} dealt with the simpler problem of  the free propagation of free massless particles in one-dimensional space. An expression similar to (\ref{eq:TrustMeMaybePlusMinusMinusMinusX}) was derived. As mentioned elsewhere in this manuscript, we intend to return to the specific features of the near-field in an upcoming work. Also, note that (\ref{eq:ConvWithDeltavpFuture}) can be rewritten
\begin{equation} \label{eq:DownOntheGround}
H_n^{\left(\pm\right)}\left(\left|\left|\mathbf{x}\right|\right|,t\right)=\frac{1}{2}\bar{f}_n\left(\mp\left|\left|\mathbf{x}\right|\right|,t\right)-C_n\left(\mp\left|\left|\mathbf{x}\right|\right|,t\right),
\end{equation}
and the photon wave function is given, according to (\ref{eq:DipoleGreen}), (\ref{eq:MyStart}) and (\ref{eq:ThisIsK}), by
\begin{multline} \label{eq:BacktoStart}
\bm{\psi}_\perp\left(\mathbf{x},t\right)=-\mathrm{i}\frac{2^{\frac{7}{2}}}{3^4}\frac{\hbar e}{\epsilon_0m_ea_0}\frac{\mathrm{e}^{-\mathrm{i}\omega_0t}}{\left(2\pi\right)^2}\\
\left[
\begin{array}{lrclclcl}
\frac{\xi_{m_2}^{\left(x,y\right)}}{\left|\left|\mathbf{x}\right|\right|}&\left[\vphantom{\left(H_0^{\left(-\right)}\right)}\right.&\left.\left[H_1^{\left(-\right)}-H_1^{\left(+\right)}\right]\left(\left|\left|\mathbf{x}\right|\right|,t\right)\right.&-&\left.\frac{\mathrm{i}}{\left|\left|\mathbf{x}\right|\right|}\left[H_2^{\left(-\right)}+H_2^{\left(+\right)}\right]\left(\left|\left|\mathbf{x}\right|\right|,t\right)\right.&-&\left.\frac{1}{\left|\left|\mathbf{x}\right|\right|^2}\left[H_3^{\left(-\right)}-H_3^{\left(+\right)}\right]\left(\left|\left|\mathbf{x}\right|\right|,t\right)\right]\\
2\frac{\xi_{m_2}^{\left(z\right)}}{\left|\left|\mathbf{x}\right|\right|}&\left[\vphantom{\left(H_0^{\left(-\right)}\right)}\right.&&&\left.\frac{\mathrm{i}}{\left|\left|\mathbf{x}\right|\right|}\left[H_2^{\left(-\right)}+H_2^{\left(+\right)}\right]\left(\left|\left|\mathbf{x}\right|\right|,t\right)\right.&+&\left.\frac{1}{\left|\left|\mathbf{x}\right|\right|^2}\left[H_3^{\left(-\right)}-H_3^{\left(+\right)}\right]\left(\left|\left|\mathbf{x}\right|\right|,t\right)\right]
\end{array}
\right]
\end{multline}
\end{widetext}
where it is of course understood that the values of all functions $H_i^{\left(\pm\right)}$ are taken at $\left(\left|\left|\mathbf{x}\right|\right|,t\right)$. Remember that $\xi_{m_2}^{\left(x,y\right)}$ are the components of $\bm{\xi}_{m_2}$ in the plane perpendicular to $\mathbf{x}$, while $\xi_{m_2}^{\left(z\right)}$ is the component of $\bm{\xi}_{m_2}$ in the direction of $\mathbf{x}$.
\begin{figure*}[t,h]
\begin{center}
\subfloat{\includegraphics[width=1.02875\columnwidth]{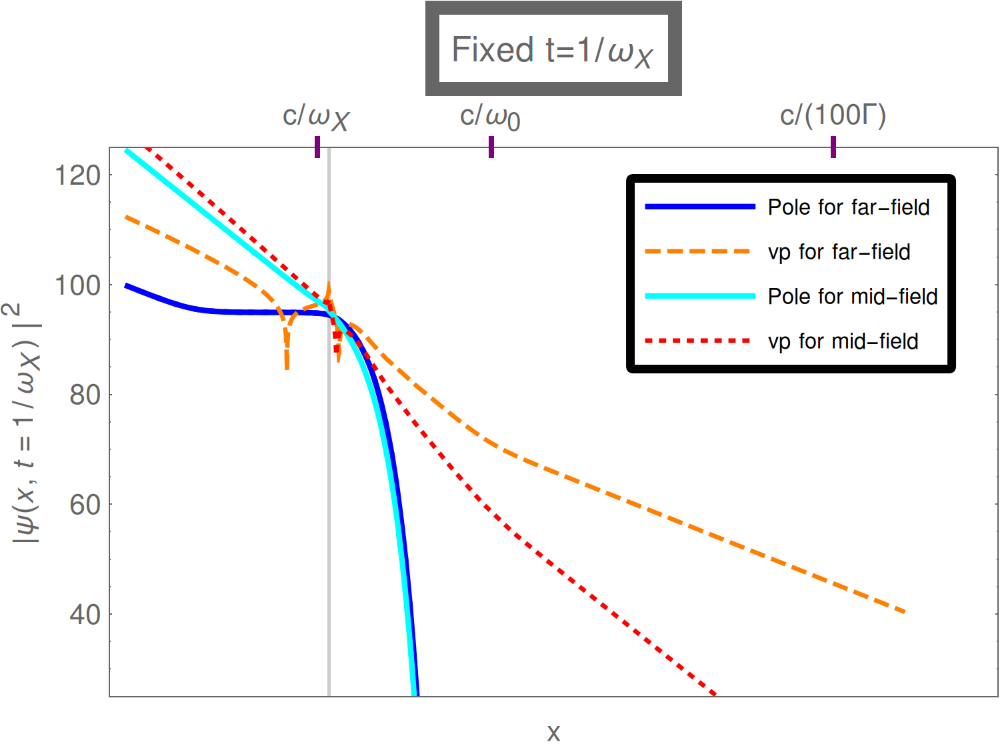}
\includegraphics[width=1.02875\columnwidth]{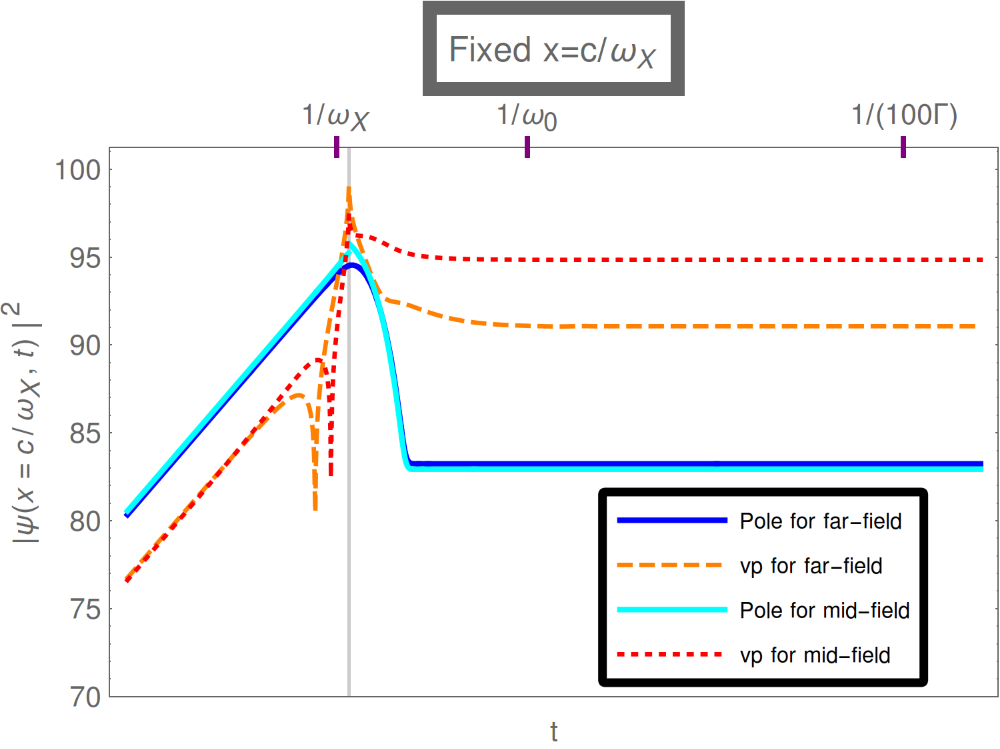}
}
\end{center}
%\caption{See caption below. \label{fig:ApMti}}
%\end{figure*}
%\begin{figure*}[t,h]
%\ContinuedFloat
\begin{center}
\subfloat{
\includegraphics[width=1.02875\columnwidth]{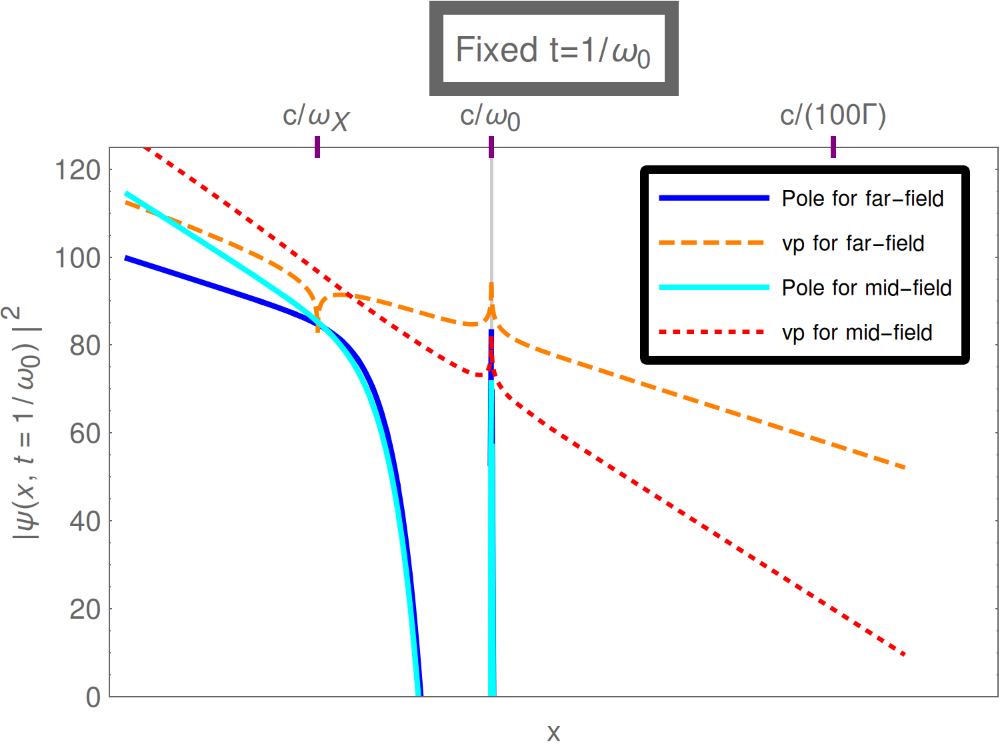}
\includegraphics[width=1.02875\columnwidth]{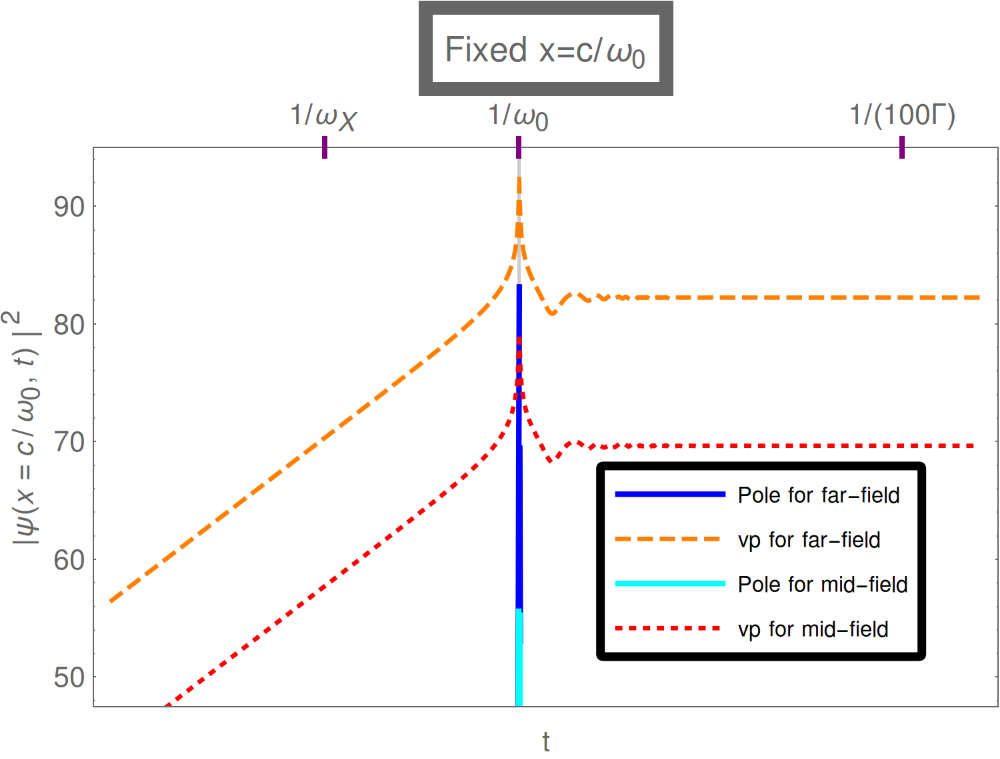}
}
\end{center}
\caption{See rest of figure and caption below. \label{fig:ApMti}}
\end{figure*}
\begin{figure*}[t,h]
\ContinuedFloat
\begin{center}
\subfloat{\includegraphics[width=1.02875\columnwidth]{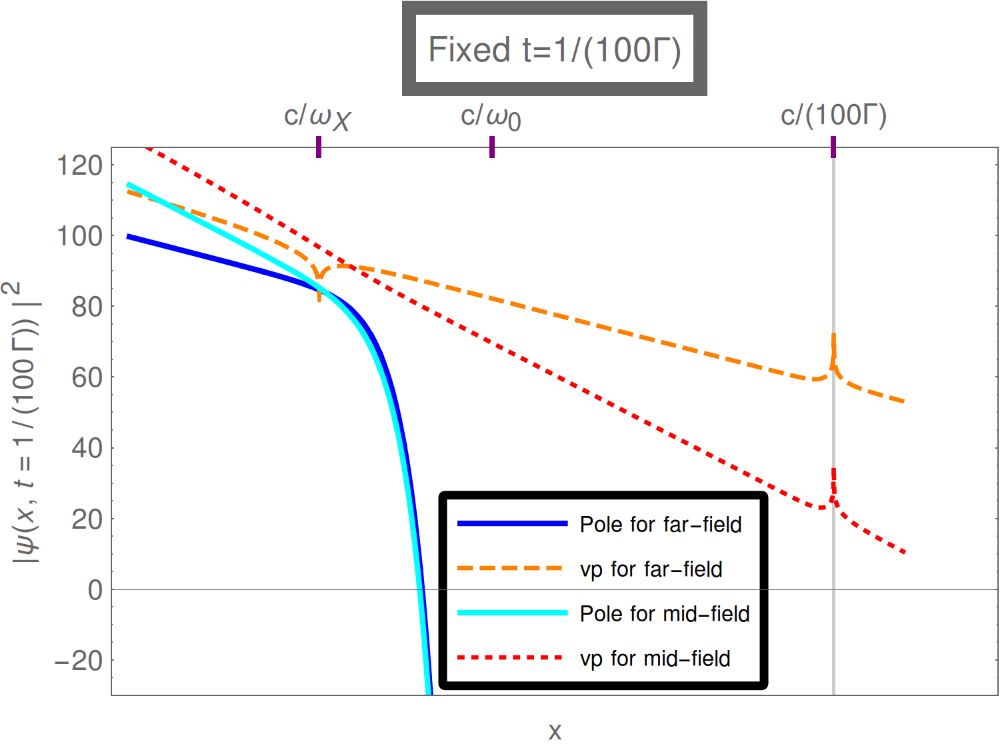}
\includegraphics[width=1.02875\columnwidth]{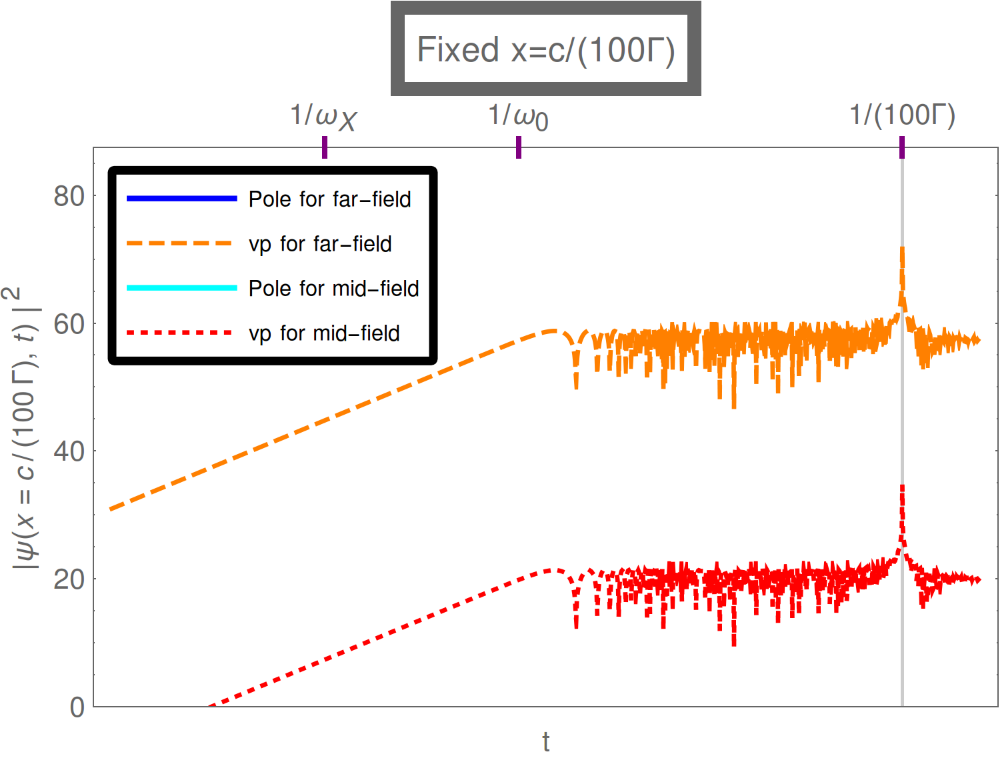}}
\end{center}
\caption{Square moduli of the contributions to the far- ($n=1$) and mid- ($n=2$) fields from the poles as given by (\ref{eq:PlusMinusMinusX}) and from the principal value integrals as given by (\ref{eq:TrustMeMaybeX}) as a function of distance from the nucleus, for fixed time in the left column, and as a function of time, for fixed distance in the right column. Only the contributions to (\ref{eq:PlusMinusMinusX}) and (\ref{eq:TrustMeMaybeX}) from the poles at $k=\pm\mathrm{i}k_{\mathrm{X}}$ are plotted here (see Fig.~\ref{fig:ApDip} for those coming from the pole $k=\omega_0/c$), which means that $G_0^{\left(n\right)}$ is artificially brought to zero. This is for the minimal $\hat{\mathbf{A}}\cdot\hat{\mathbf{p}}$ coupling (sect.~\ref{sec:MyHeart}). Both axes are logarithmic on all figures. The boundary of the lightcone is signalled by a vertical line. \label{fig:ApMti}}
\end{figure*}

\section{Discussion} \label{sec:Dsc}

According to (\ref{eq:CplxResidue}) and (\ref{eq:TrustMeMaybeX})\textemdash and, in the $n=3$ case, (\ref{eq:CplxResidueZeroX}) and (\ref{eq:TrustMeMaybeZeroX})\textemdash our final result given by (\ref{eq:DownOntheGround}) and (\ref{eq:BacktoStart}) is evidently noncausal. We identify three sources of noncausality:
\begin{itemize}
\item{The first one is well-known \cite{Shirokov,HegerfeldtFermi} and is due to the fact that, in order to compute the photon wave function, we integrated over the physical frequencies of the electromagnetic field, which are positive. Indeed, the spectrum of the electromagnetic field Hamiltonian is the positive real semi-axis and is thus bounded from below. In light of Hegerfeldt's theorem, it is then not surprising to observe that our result is noncausal. More precisely, the theorem teaches that the convolution contributions (\ref{eq:TrustMeMaybeX}) and (\ref{eq:TrustMeMaybeZeroX}), which would not be featured if the integration was carried out over the whole real axis, necessarily introduce a noncausality in (\ref{eq:DownOntheGround}). This feature has already been studied in sect.~\ref{subsec:AdotPDipHeg}, where the dipole approximation was performed, which changes little to the discussion and the results.}
\item{The second one is the presence of the singularity at $k=0$ of the integrand (\ref{eq:ThisIsK}) for $n=3$, that is, according to (\ref{eq:DipoleGreen}), in the near-field zone of emission. Terms coming from this singularity are given in (\ref{eq:CplxResidueZeroX}) and (\ref{eq:TrustMeMaybeZeroX}). As can be seen from (\ref{eq:BacktoStart}), this part of the emitted field decays as the third power $\left|\left|\mathbf{x}\right|\right|^{-3}$ of the inverse distance from the nucleus to the point of observation. There is much of interest to say about the near-field and this noncausality, and how they can be linked to the (Coulomb) longitudinal electric field, but we will discuss that in a different manuscript.}
\item{The third and last one is the uncertainty on the position of the electron. For instance, at $t=0$, when the emission starts, the electron is in the $2\mathrm{p}$ level, and its wave function is spread on a distance of order $2a_0$. In the light of this uncertainty, it is natural to expect a spacewise-exponentially decreasing tail in the emitted field, with characteristic size of order $a_0$. And this is what we indeed obtain if we neglect the other sources of noncausality, namely if
\begin{itemize}
\item{we integrate in (\ref{eq:MyStart}) over both positive and negative frequencies,}
\item{we only focus on the mid- ($n=2$) and far- ($n=1$) field zones of emission,}
\end{itemize}
in other words, if we consider only expression (\ref{eq:CplxResidue}), the result is still noncausal. In the previous sentence ``noncausal'' is understood to mean ``not vanishing outside the lightcone centred around $t=0$ and the position $\mathbf{x}=\mathbf{0}$ of the Hydrogen nucleus (proton)''. Rather, the noncausal terms in (\ref{eq:CplxResidue}) decay exponentially on a distance $3a_0/2$ (see Fig.~\ref{fig:ApMti}). Hence these terms are nonvanishing outside the lightcone centred around $t=0$ and $\mathbf{x}=\mathbf{0}$, which we understand as being an illustration of the fact that the electron emits the photon from its own position, which is not fully determined and is only exponentially confined within distances of order $a_0$ around the nucleus, rather than from the position $\mathbf{x}=0$ of the nucleus itself \footnote{It is only if the dipole approximation is performed that the emission can be considered to take place at $\mathbf{x}=\mathbf{0}$. As discussed at length in \cite{EdouardIsa} and \cite{FredZ}, and also below (\ref{eq:ExpInt}), the dipole approximation results in divergences coming from ultraviolet frequencies. A possibility is then to implement a cutoff on high frequencies (the cutoff being of order $c/a_0$) which, of course, will induce a ``blurring'' of the lightcone over distances of order $a_0$, in virtue of the Paley-Wiener properties of the Fourier transform. In this sense, the noncausality due to the finite size of the decaying electron, first noticed by Shirokov \cite{Shirokov} and described here in detail, is similar to Hergerfeldt's noncausality.}. In \cite{Shirokov}, Shirokov very clearly hints at such exponentially decreasing tails outside the lightcone. But among the works of his which are available to us (including \cite{Vlad}), none presents or even mentions the method he used to obtain this feature. Note that the contributions to the emitted field from the poles at $k=\pm\mathrm{i}k_{\mathrm{X}}$ (which correspond to (\ref{eq:PlusMinusMinusX}) and (\ref{eq:TrustMeMaybeX}) with $G_0^{\left(n\right)}$ artificially brought to zero)  which arise when the dipole approximation is not performed, are also subject, so to say, to Hegerfeldt noncausality. This results in the (principal value) noncausal contributions seen in Fig.~\ref{fig:ApMti}. Interestingly, these noncausal terms do not decay more strongly with increasing distance outside the lightcone than they do inside, as was the case for the noncausal terms studied in sect.~\ref{sec:PreStorm}. This can be seen either graphically or by noticing that the same asymptotic arguments as developed in sect.~\ref{sec:PreStorm} do not yield the same result here.}
\end{itemize}
In an upcoming manuscript we will focus particularly on the near-field, where the longitudinal (in the sense of Fourier space, see \cite{CohenQED1}) contribution to the electric field comes into play.\\

The results presented here regarding departure from causality are, of course, disturbing but maybe not exceedingly disturbing. Indeed, one notices from Figs.~\ref{fig:ApDip} and \ref{fig:ApMti} that the weight of the wave function present outside the light cone is always small. In \cite{DDDNVZHaroche} for instance we estimated that for a simpler but related problem, the weight of the negative frequencies after a time of the order of the lifetime is bounded by $10^{-10}$, which could never explain for instance non-locality effects à la Bell characterized by violations of $1/\sqrt{2}$ (see footnote 3 in \cite{DDDNVZHaroche}).\\

Moreover Sipe showed \cite{Sipe} that, performing the same kind of approximations as described in sect.~\ref{sec:CausalWW} (usual $\hat{\mathbf{E}}\cdot\hat{\mathbf{x}}$ coupling in the dipole approximation, integration over the whole real axis), one can approximate QED by a theory in which the photon wave function obeys (complex) Maxwell-type equations and thus admits a causal (retarded) Green's function. This explains why departures from causality are always small. It is not clear at this level whether QED is intrinsically non-causal or whether apparent non-causalities result from approximations performed somewhere in the theoretical developments.

\begin{acknowledgments}
Vincent Debierre acknowledges support from CNRS (INSIS doctoral grant). Thomas Durt acknowledges support from the COST 1006 and COST 1403 actions. We thank Pr. Édouard Brainis for helpful discussions. We also thank Nikolai Korchagin at the Joint Institute for Nuclear Research in Dubna for sending us a copy of Shirokov's paper \cite{Vlad}. Very warm thanks go to Vladyslav Atavin who sat down with one of us (VD) to patiently translate Shirokov's paper \cite{Vlad} from Russian.
\end{acknowledgments}

\bibliography{Biblio}

\begin{thebibliography}{30}%
\makeatletter
\providecommand \@ifxundefined [1]{%
 \@ifx{#1\undefined}
}%
\providecommand \@ifnum [1]{%
 \ifnum #1\expandafter \@firstoftwo
 \else \expandafter \@secondoftwo
 \fi
}%
\providecommand \@ifx [1]{%
 \ifx #1\expandafter \@firstoftwo
 \else \expandafter \@secondoftwo
 \fi
}%
\providecommand \natexlab [1]{#1}%
\providecommand \enquote  [1]{``#1''}%
\providecommand \bibnamefont  [1]{#1}%
\providecommand \bibfnamefont [1]{#1}%
\providecommand \citenamefont [1]{#1}%
\providecommand \href@noop [0]{\@secondoftwo}%
\providecommand \href [0]{\begingroup \@sanitize@url \@href}%
\providecommand \@href[1]{\@@startlink{#1}\@@href}%
\providecommand \@@href[1]{\endgroup#1\@@endlink}%
\providecommand \@sanitize@url [0]{\catcode `\\12\catcode `\$12\catcode
  `\&12\catcode `\#12\catcode `\^12\catcode `\_12\catcode `\%12\relax}%
\providecommand \@@startlink[1]{}%
\providecommand \@@endlink[0]{}%
\providecommand \url  [0]{\begingroup\@sanitize@url \@url }%
\providecommand \@url [1]{\endgroup\@href {#1}{\urlprefix }}%
\providecommand \urlprefix  [0]{URL }%
\providecommand \Eprint [0]{\href }%
\providecommand \doibase [0]{http://dx.doi.org/}%
\providecommand \selectlanguage [0]{\@gobble}%
\providecommand \bibinfo  [0]{\@secondoftwo}%
\providecommand \bibfield  [0]{\@secondoftwo}%
\providecommand \translation [1]{[#1]}%
\providecommand \BibitemOpen [0]{}%
\providecommand \bibitemStop [0]{}%
\providecommand \bibitemNoStop [0]{.\EOS\space}%
\providecommand \EOS [0]{\spacefactor3000\relax}%
\providecommand \BibitemShut  [1]{\csname bibitem#1\endcsname}%
\let\auto@bib@innerbib\@empty
%</preamble>
\bibitem [{\citenamefont {Fermi}(1932)}]{FermiCausal}%
  \BibitemOpen
  \bibfield  {author} {\bibinfo {author} {\bibfnamefont {E.}~\bibnamefont
  {Fermi}},\ }\href {http://rmp.aps.org/abstract/RMP/v4/i1/p87_1} {\bibfield
  {journal} {\bibinfo  {journal} {Rev. Mod. Phys.}\ }\textbf {\bibinfo {volume}
  {4}},\ \bibinfo {pages} {87} (\bibinfo {year} {1932})}\BibitemShut {NoStop}%
\bibitem [{\citenamefont {Shirokov}(1978)}]{Shirokov}%
  \BibitemOpen
  \bibfield  {author} {\bibinfo {author} {\bibfnamefont {M.}~\bibnamefont
  {Shirokov}},\ }\href {http://iopscience.iop.org/0038-5670/21/4/R05}
  {\bibfield  {journal} {\bibinfo  {journal} {Sov. Phys. Usp.}\ }\textbf
  {\bibinfo {volume} {21}},\ \bibinfo {pages} {345} (\bibinfo {year}
  {1978})}\BibitemShut {NoStop}%
\bibitem [{\citenamefont {Biswas}\ \emph {et~al.}(1990)\citenamefont {Biswas},
  \citenamefont {Compagno}, \citenamefont {Palma}, \citenamefont {Passante},\
  and\ \citenamefont {Persico}}]{PassanteVirtual}%
  \BibitemOpen
  \bibfield  {author} {\bibinfo {author} {\bibfnamefont {A.}~\bibnamefont
  {Biswas}}, \bibinfo {author} {\bibfnamefont {G.}~\bibnamefont {Compagno}},
  \bibinfo {author} {\bibfnamefont {G.}~\bibnamefont {Palma}}, \bibinfo
  {author} {\bibfnamefont {R.}~\bibnamefont {Passante}}, \ and\ \bibinfo
  {author} {\bibfnamefont {F.}~\bibnamefont {Persico}},\ }\href
  {http://journals.aps.org/pra/abstract/10.1103/PhysRevA.42.4291} {\bibfield
  {journal} {\bibinfo  {journal} {Phys. Rev. A}\ }\textbf {\bibinfo {volume}
  {42}},\ \bibinfo {pages} {4291} (\bibinfo {year} {1990})}\BibitemShut
  {NoStop}%
\bibitem [{\citenamefont {Hegerfeldt}(1994)}]{HegerfeldtFermi}%
  \BibitemOpen
  \bibfield  {author} {\bibinfo {author} {\bibfnamefont {G.}~\bibnamefont
  {Hegerfeldt}},\ }\href
  {http://journals.aps.org/prl/abstract/10.1103/PhysRevLett.72.596} {\bibfield
  {journal} {\bibinfo  {journal} {Phys. Rev. Lett.}\ }\textbf {\bibinfo
  {volume} {72}},\ \bibinfo {pages} {596} (\bibinfo {year} {1994})}\BibitemShut
  {NoStop}%
\bibitem [{\citenamefont {Sipe}(1995)}]{Sipe}%
  \BibitemOpen
  \bibfield  {author} {\bibinfo {author} {\bibfnamefont {J.}~\bibnamefont
  {Sipe}},\ }\href {http://pra.aps.org/abstract/PRA/v52/i3/p1875_1} {\bibfield
  {journal} {\bibinfo  {journal} {Phys. Rev. A}\ }\textbf {\bibinfo {volume}
  {52}},\ \bibinfo {pages} {1875} (\bibinfo {year} {1995})}\BibitemShut
  {NoStop}%
\bibitem [{\citenamefont {Weisskopf}\ and\ \citenamefont {Wigner}(1930)}]{WW}%
  \BibitemOpen
  \bibfield  {author} {\bibinfo {author} {\bibfnamefont {V.}~\bibnamefont
  {Weisskopf}}\ and\ \bibinfo {author} {\bibfnamefont {E.}~\bibnamefont
  {Wigner}},\ }\href {http://link.springer.com/article/10.1007%2FBF01336768}
  {\bibfield  {journal} {\bibinfo  {journal} {Z. Phys.}\ }\textbf {\bibinfo
  {volume} {63}},\ \bibinfo {pages} {54} (\bibinfo {year} {1930})}\BibitemShut
  {NoStop}%
\bibitem [{\citenamefont {Paley}\ and\ \citenamefont
  {Wiener}(1934)}]{PaleyWiener}%
  \BibitemOpen
  \bibfield  {author} {\bibinfo {author} {\bibfnamefont {R.}~\bibnamefont
  {Paley}}\ and\ \bibinfo {author} {\bibfnamefont {N.}~\bibnamefont {Wiener}},\
  }\href@noop {} {\emph {\bibinfo {title} {Fourier Transforms In the Complex
  Domain}}}\ (\bibinfo  {publisher} {The American Mathematical Society},\
  \bibinfo {year} {1934})\BibitemShut {NoStop}%
\bibitem [{\citenamefont {Seke}(1994)}]{Seke}%
  \BibitemOpen
  \bibfield  {author} {\bibinfo {author} {\bibfnamefont {J.}~\bibnamefont
  {Seke}},\ }\href
  {http://www.sciencedirect.com/science/article/pii/0378437194901562}
  {\bibfield  {journal} {\bibinfo  {journal} {Phys. A}\ }\textbf {\bibinfo
  {volume} {203}},\ \bibinfo {pages} {269} (\bibinfo {year}
  {1994})}\BibitemShut {NoStop}%
\bibitem [{\citenamefont {Facchi}(2000)}]{FacchiPhD}%
  \BibitemOpen
  \bibfield  {author} {\bibinfo {author} {\bibfnamefont {P.}~\bibnamefont
  {Facchi}},\ }\href {http://www.ba.infn.it/~facchi/lectures/thesis.pdf}
  {\enquote {\bibinfo {title} {Quantum time evolution: Free and controlled
  dynamics},}\ } (\bibinfo {year} {2000})\BibitemShut {NoStop}%
\bibitem [{\citenamefont {Scully}\ and\ \citenamefont
  {Zubairy}(1997)}]{FoxMulder}%
  \BibitemOpen
  \bibfield  {author} {\bibinfo {author} {\bibfnamefont {M.}~\bibnamefont
  {Scully}}\ and\ \bibinfo {author} {\bibfnamefont {M.}~\bibnamefont
  {Zubairy}},\ }\href@noop {} {\emph {\bibinfo {title} {Quantum Optics}}}\
  (\bibinfo  {publisher} {Cambridge University Press},\ \bibinfo {year}
  {1997})\BibitemShut {NoStop}%
\bibitem [{\citenamefont {Cohen-Tannoudji}\ \emph {et~al.}(2001)\citenamefont
  {Cohen-Tannoudji}, \citenamefont {Dupont-Roc},\ and\ \citenamefont
  {Grynberg}}]{CohenQED1}%
  \BibitemOpen
  \bibfield  {author} {\bibinfo {author} {\bibfnamefont {C.}~\bibnamefont
  {Cohen-Tannoudji}}, \bibinfo {author} {\bibfnamefont {J.}~\bibnamefont
  {Dupont-Roc}}, \ and\ \bibinfo {author} {\bibfnamefont {G.}~\bibnamefont
  {Grynberg}},\ }\href@noop {} {\emph {\bibinfo {title} {Photons et atomes -
  Introduction à l'électrodynamique quantique}}},\ \bibinfo {edition} {2nd}\
  ed.\ (\bibinfo  {publisher} {EDP Sciences/CNRS Éditions},\ \bibinfo {year}
  {2001})\BibitemShut {NoStop}%
\bibitem [{\citenamefont {Weinberg}(1995)}]{Weinberg1}%
  \BibitemOpen
  \bibfield  {author} {\bibinfo {author} {\bibfnamefont {S.}~\bibnamefont
  {Weinberg}},\ }\href@noop {} {\emph {\bibinfo {title} {The Quantum Theory of
  Fields}}},\ \bibinfo {edition} {1st}\ ed.,\ Vol.~\bibinfo {volume} {1}\
  (\bibinfo  {publisher} {Cambridge University Press},\ \bibinfo {year}
  {1995})\BibitemShut {NoStop}%
\bibitem [{\citenamefont {Itzykson}\ and\ \citenamefont
  {Zuber}(1980)}]{ItzyksonZuber}%
  \BibitemOpen
  \bibfield  {author} {\bibinfo {author} {\bibfnamefont {C.}~\bibnamefont
  {Itzykson}}\ and\ \bibinfo {author} {\bibfnamefont {J.}~\bibnamefont
  {Zuber}},\ }\href@noop {} {\emph {\bibinfo {title} {Quantum Field Theory}}},\
  \bibinfo {edition} {1st}\ ed.\ (\bibinfo  {publisher} {McGraw-Hill},\
  \bibinfo {year} {1980})\BibitemShut {NoStop}%
\bibitem [{\citenamefont {Debierre}\ \emph
  {et~al.}(2015{\natexlab{a}})\citenamefont {Debierre}, \citenamefont
  {Goessens}, \citenamefont {Brainis},\ and\ \citenamefont
  {Durt}}]{EdouardIsa}%
  \BibitemOpen
  \bibfield  {author} {\bibinfo {author} {\bibfnamefont {V.}~\bibnamefont
  {Debierre}}, \bibinfo {author} {\bibfnamefont {I.}~\bibnamefont {Goessens}},
  \bibinfo {author} {\bibfnamefont {E.}~\bibnamefont {Brainis}}, \ and\
  \bibinfo {author} {\bibfnamefont {T.}~\bibnamefont {Durt}},\ }\href
  {http://journals.aps.org/pra/abstract/10.1103/PhysRevA.92.023825} {\bibfield
  {journal} {\bibinfo  {journal} {Phys. Rev. A}\ }\textbf {\bibinfo {volume}
  {92}},\ \bibinfo {pages} {023825} (\bibinfo {year}
  {2015}{\natexlab{a}})}\BibitemShut {NoStop}%
\bibitem [{\citenamefont {Akhiezer}\ and\ \citenamefont
  {Berestetskii}(1965)}]{TellementOuf}%
  \BibitemOpen
  \bibfield  {author} {\bibinfo {author} {\bibfnamefont {A.}~\bibnamefont
  {Akhiezer}}\ and\ \bibinfo {author} {\bibfnamefont {V.}~\bibnamefont
  {Berestetskii}},\ }\href@noop {} {\emph {\bibinfo {title} {Quantum
  Electrodynamics}}},\ \bibinfo {edition} {2nd}\ ed.,\ edited by\ \bibinfo
  {editor} {\bibfnamefont {R.}~\bibnamefont {Marshak}}\ (\bibinfo  {publisher}
  {Interscience Publishers},\ \bibinfo {year} {1965})\BibitemShut {NoStop}%
\bibitem [{\citenamefont {Białynicki-Birula}(1996)}]{BBPWF}%
  \BibitemOpen
  \bibfield  {author} {\bibinfo {author} {\bibfnamefont {I.}~\bibnamefont
  {Białynicki-Birula}},\ }\href
  {http://www.sciencedirect.com/science/article/pii/S0079663808703160}
  {\bibfield  {journal} {\bibinfo  {journal} {Prog. Opt.}\ }\textbf {\bibinfo
  {volume} {36}},\ \bibinfo {pages} {245} (\bibinfo {year} {1996})}\BibitemShut
  {NoStop}%
\bibitem [{\citenamefont {Hawton}(2008)}]{HawtonLorentz}%
  \BibitemOpen
  \bibfield  {author} {\bibinfo {author} {\bibfnamefont {M.}~\bibnamefont
  {Hawton}},\ }\href {http://pra.aps.org/abstract/PRA/v78/i1/e012111}
  {\bibfield  {journal} {\bibinfo  {journal} {Phys. Rev. A}\ }\textbf {\bibinfo
  {volume} {78}},\ \bibinfo {pages} {012111} (\bibinfo {year}
  {2008})}\BibitemShut {NoStop}%
\bibitem [{\citenamefont {Smith}\ and\ \citenamefont
  {Raymer}(2007)}]{RaymerSmith}%
  \BibitemOpen
  \bibfield  {author} {\bibinfo {author} {\bibfnamefont {B.}~\bibnamefont
  {Smith}}\ and\ \bibinfo {author} {\bibfnamefont {M.}~\bibnamefont {Raymer}},\
  }\href {http://iopscience.iop.org/1367-2630/9/11/414} {\bibfield  {journal}
  {\bibinfo  {journal} {New J. Phys.}\ }\textbf {\bibinfo {volume} {9}},\
  \bibinfo {pages} {414} (\bibinfo {year} {2007})}\BibitemShut {NoStop}%
\bibitem [{\citenamefont {Titulaer}\ and\ \citenamefont
  {Glauber}(1966)}]{Titulaer}%
  \BibitemOpen
  \bibfield  {author} {\bibinfo {author} {\bibfnamefont {U.}~\bibnamefont
  {Titulaer}}\ and\ \bibinfo {author} {\bibfnamefont {R.}~\bibnamefont
  {Glauber}},\ }\href
  {http://journals.aps.org/pr/abstract/10.1103/PhysRev.145.1041} {\bibfield
  {journal} {\bibinfo  {journal} {Phys. Rev.}\ }\textbf {\bibinfo {volume}
  {145}},\ \bibinfo {pages} {1041} (\bibinfo {year} {1966})}\BibitemShut
  {NoStop}%
\bibitem [{\citenamefont {Facchi}\ and\ \citenamefont
  {Pascazio}(1998)}]{FacchiPascazio}%
  \BibitemOpen
  \bibfield  {author} {\bibinfo {author} {\bibfnamefont {P.}~\bibnamefont
  {Facchi}}\ and\ \bibinfo {author} {\bibfnamefont {S.}~\bibnamefont
  {Pascazio}},\ }\href
  {http://www.sciencedirect.com/science/article/pii/S0375960198001443}
  {\bibfield  {journal} {\bibinfo  {journal} {Phys. Lett. A}\ }\textbf
  {\bibinfo {volume} {241}},\ \bibinfo {pages} {139} (\bibinfo {year}
  {1998})}\BibitemShut {NoStop}%
\bibitem [{\citenamefont {Appel}(2007)}]{Appel}%
  \BibitemOpen
  \bibfield  {author} {\bibinfo {author} {\bibfnamefont {W.}~\bibnamefont
  {Appel}},\ }\href@noop {} {\emph {\bibinfo {title} {Mathematics for Physics
  \& Physicists}}},\ \bibinfo {edition} {1st}\ ed.\ (\bibinfo  {publisher}
  {Princeton University Press},\ \bibinfo {year} {2007})\BibitemShut {NoStop}%
\bibitem [{\citenamefont {Debierre}\ \emph
  {et~al.}(2015{\natexlab{b}})\citenamefont {Debierre}, \citenamefont {Durt},
  \citenamefont {Nicolet},\ and\ \citenamefont {Zolla}}]{FredZ}%
  \BibitemOpen
  \bibfield  {author} {\bibinfo {author} {\bibfnamefont {V.}~\bibnamefont
  {Debierre}}, \bibinfo {author} {\bibfnamefont {T.}~\bibnamefont {Durt}},
  \bibinfo {author} {\bibfnamefont {A.}~\bibnamefont {Nicolet}}, \ and\
  \bibinfo {author} {\bibfnamefont {F.}~\bibnamefont {Zolla}},\ }\href
  {http://www.sciencedirect.com/science/article/pii/S0375960115005307}
  {\bibfield  {journal} {\bibinfo  {journal} {accepted by Phys. Lett. A}\ }
  (\bibinfo {year} {2015}{\natexlab{b}})}\BibitemShut {NoStop}%
\bibitem [{Note1()}]{Note1}%
  \BibitemOpen
  \bibinfo {note} {Let us note that, for $n\in \left \protect \{1,2\right
  \protect \}$, the integrand in (\ref {eq:Bob}) actually is an entire function
  of $k$, since the singularity at $\Omega _0/c$ is also artificial. To compute
  the integral, though, we split the integrand into two meromorphic functions
  in order to be able to use the Jordan lemma.}\BibitemShut {Stop}%
\bibitem [{\citenamefont {Abramowitz}\ and\ \citenamefont
  {Stegun}(1964)}]{AbramowitzStegun}%
  \BibitemOpen
  \bibfield  {author} {\bibinfo {author} {\bibfnamefont {M.}~\bibnamefont
  {Abramowitz}}\ and\ \bibinfo {author} {\bibfnamefont {I.}~\bibnamefont
  {Stegun}},\ }\href@noop {} {\emph {\bibinfo {title} {Handbook of Mathematical
  Functions with Formulas, Graphs, and Mathematical Tables}}}\ (\bibinfo
  {publisher} {Dover},\ \bibinfo {year} {1964})\BibitemShut {NoStop}%
\bibitem [{Note2()}]{Note2}%
  \BibitemOpen
  \bibinfo {note} {This was also true of the integrals $H_{n\left (\protect
  \mathrm {std}\right )}^{\left (\pm \right )}$ for $n=1,2$ in the $\protect
  \mathaccentV {hat}05E{\protect \mathbf {E}}\cdot \protect \mathaccentV
  {hat}05E{\protect \mathbf {x}}$ case of the previous sect.~\ref
  {sec:CausalWW}.}\BibitemShut {Stop}%
\bibitem [{Note3()}]{Note3}%
  \BibitemOpen
  \bibinfo {note} {Note that it is very easy to show that if, instead of this
  time-dependent perturbation theory, we plug in the expression from the usual
  Wigner-Weisskopf approximation (which has been shown \cite {FacchiPascazio}
  to provide a very good approximation to the dynamics of the system (except at
  very short times), up to very small corrections), our results for the matrix
  elements of the electric field hold, with a grain of salt: $\omega _0$ should
  be replaced by $\omega _0+\omega _{\protect \mathrm {LS}}-\left (\protect
  \mathrm {i}/2\right )\Gamma $. Therefore, our perturbative treatment works
  directly at short times and indirectly (through the substitution which we
  just specified) at intermediate times. It only fails at very long times,
  where the decay becomes nonexponential \cite {FacchiPascazio}.}\BibitemShut
  {Stop}%
\bibitem [{\citenamefont {Karpov}\ \emph {et~al.}(2000)\citenamefont {Karpov},
  \citenamefont {Ordonez}, \citenamefont {Petrosky}, \citenamefont {Progine},\
  and\ \citenamefont {Pronko}}]{KarpovCausal}%
  \BibitemOpen
  \bibfield  {author} {\bibinfo {author} {\bibfnamefont {E.}~\bibnamefont
  {Karpov}}, \bibinfo {author} {\bibfnamefont {G.}~\bibnamefont {Ordonez}},
  \bibinfo {author} {\bibfnamefont {T.}~\bibnamefont {Petrosky}}, \bibinfo
  {author} {\bibfnamefont {I.}~\bibnamefont {Progine}}, \ and\ \bibinfo
  {author} {\bibfnamefont {G.}~\bibnamefont {Pronko}},\ }\href
  {http://journals.aps.org/pra/abstract/10.1103/PhysRevA.62.012103} {\bibfield
  {journal} {\bibinfo  {journal} {Phys. Rev. A}\ }\textbf {\bibinfo {volume}
  {62}},\ \bibinfo {pages} {012103} (\bibinfo {year} {2000})}\BibitemShut
  {NoStop}%
\bibitem [{Note4()}]{Note4}%
  \BibitemOpen
  \bibinfo {note} {It is only if the dipole approximation is performed that the
  emission can be considered to take place at $\protect \mathbf {x}=\protect
  \mathbf {0}$. As discussed at length in \cite {EdouardIsa} and \cite {FredZ},
  and also below (\ref {eq:ExpInt}), the dipole approximation results in
  divergences coming from ultraviolet frequencies. A possibility is then to
  implement a cutoff on high frequencies (the cutoff being of order $c/a_0$)
  which, of course, will induce a ``blurring'' of the lightcone over distances
  of order $a_0$, in virtue of the Paley-Wiener properties of the Fourier
  transform. In this sense, the noncausality due to the finite size of the
  decaying electron, first noticed by Shirokov \cite {Shirokov} and described
  here in detail, is similar to Hergerfeldt's noncausality.}\BibitemShut
  {Stop}%
\bibitem [{\citenamefont {Shirokov}(1964)}]{Vlad}%
  \BibitemOpen
  \bibfield  {author} {\bibinfo {author} {\bibfnamefont {M.}~\bibnamefont
  {Shirokov}},\ }\href {http://www.osti.gov/scitech/biblio/4009861} {\bibfield
  {journal} {\bibinfo  {journal} {JINR Preprint}\ ,\ \bibinfo {pages} {1719}}
  (\bibinfo {year} {1964})}\BibitemShut {NoStop}%
\bibitem [{\citenamefont {Debierre}\ \emph {et~al.}(2014)\citenamefont
  {Debierre}, \citenamefont {Demésy}, \citenamefont {Durt}, \citenamefont
  {Nicolet}, \citenamefont {Vial},\ and\ \citenamefont
  {Zolla}}]{DDDNVZHaroche}%
  \BibitemOpen
  \bibfield  {author} {\bibinfo {author} {\bibfnamefont {V.}~\bibnamefont
  {Debierre}}, \bibinfo {author} {\bibfnamefont {G.}~\bibnamefont {Demésy}},
  \bibinfo {author} {\bibfnamefont {T.}~\bibnamefont {Durt}}, \bibinfo {author}
  {\bibfnamefont {A.}~\bibnamefont {Nicolet}}, \bibinfo {author} {\bibfnamefont
  {B.}~\bibnamefont {Vial}}, \ and\ \bibinfo {author} {\bibfnamefont
  {F.}~\bibnamefont {Zolla}},\ }\href
  {http://journals.aps.org/pra/abstract/10.1103/PhysRevA.90.033806} {\bibfield
  {journal} {\bibinfo  {journal} {Phys. Rev. A}\ }\textbf {\bibinfo {volume}
  {90}},\ \bibinfo {pages} {033806} (\bibinfo {year} {2014})}\BibitemShut
  {NoStop}%
\end{thebibliography}%
\end{document}